\documentclass[pdflatex,sn-mathphys-num]{sn-jnl}%

\usepackage{graphicx}%
\usepackage{multirow}%
\usepackage{amsmath,amssymb,amsfonts}%
\usepackage{amsthm}%
\usepackage{mathrsfs}%
\usepackage[title]{appendix}%
\usepackage{xcolor}%
\usepackage{textcomp}%
\usepackage{manyfoot}%
\usepackage{booktabs}%
\usepackage{algorithm}%
\usepackage{algorithmicx}%
\usepackage{algpseudocode}%
\usepackage{listings}%
\usepackage{calc}%
\usepackage{tipa}%
\usepackage{enumitem}%
\usepackage{multicol}%
\usepackage{arydshln}%
\usepackage{colortbl}%
\usepackage{placeins}%
\theoremstyle{thmstyleone}%

\theoremstyle{thmstyletwo}%

\theoremstyle{thmstylethree}%
\newtheorem{definition}{Definition}%

\newtheorem{fact}{Fact}

\newcommand{\tico}{TSP-T3CO~}%

\newcommand{\ie}{i.e.,~}
\newcommand{\eg}{e.g.,~}
\newcommand{\sol}{\mathcal{S}}
\newcommand{\compl}[1]{\overline{#1}}%
\newcommand{\cost}{c}
\newcommand{\greyed}{\cellcolor{gray!50}}
\newcommand{\dotlinea}{\cdashline{2-6}[0.5pt/2pt]}
\newcommand{\dotlineb}{\cdashline{2-7}[0.5pt/2pt]}
\newcommand{\dotlinec}{\cdashline{2-5}[0.5pt/2pt]}
\newcommand{\confirmed}{\checkmark}  %
\newcommand{\unconfirmed}{\phantom{\confirmed}}  %
\newsavebox\CBox
\newcommand\hcancel[2][0.5pt]{%
  \ifmmode\sbox\CBox{$#2$}\else\sbox\CBox{#2}\fi%
  \makebox[0pt][l]{\usebox\CBox}%
  \rule[0.5\ht\CBox-#1/2]{\wd\CBox}{#1}}

\begin{document}

\title[TSP Approximability Results leveraging the TSP-T3CO Definition Scheme]{A Survey on Approximability of Traveling Salesman Problems using the TSP-T3CO Definition Scheme}

\author[1]{\fnm{Sophia} \sur{Saller}}
\email{ssaller@xpect.ai}
\affil[1]{%
 \orgname{Deutsches Forschungszentrum f\"ur K\"unstliche Intelligenz}%
}
\author[2]{Jana Koehler}
\email{jana.koehler@hslu.ch}
\affil[3]{%
  \orgname{Lucerne University of Applied Sciences and Arts}%
}
\author[3]{Andreas Karrenbauer}
\email{andreas.karrenbauer@mpi-inf.mpg.de}
\affil[3]{%
  \orgname{Max Planck Institute for Informatics}%
}

\abstract{The traveling salesman (or salesperson) problem, short TSP, is of strong interest to many researchers from mathematics, economics, and computer science. Manifold TSP variants occur in nearly every scientific field and application domain: \eg engineering, physics, biology, life sciences, and manufacturing. Several thousand papers are published every year. 

This paper provides the first systematic survey on the best currently known approximability and inapproximability results for well-known TSP variants such as the "standard",  Path, Bottleneck, Maximum Scatter, Generalized, Clustered, Quota, Prize-Collecting, Time-dependent TSP, Traveling Purchaser Problem, Profitable Tour Problem, Orienteering Problem,  TSP with Time Windows, and Orienteering Problem with Time Windows. 
 
The foundation of our survey is the definition scheme \tico, which we propose as a uniform, easy-to-use and extensible means for the formal and precise definition of TSP variants. Applying \tico to  define a TSP variant reveals subtle differences within the same named variant and also brings out the differences between variants more clearly. We achieve the first comprehensive, concise, and compact representation of approximability results by using \tico definitions. This makes it easier to understand the approximability landscape and the assumptions under which certain results hold. Open gaps become more evident and results can be compared more easily. }


\maketitle

\section{Introduction}

The traveling salesman problem, the gender-neutral name being the traveling salesperson problem, short TSP, is a problem of strong interest to many researchers and practitioners in the fields of mathematics, economics, computer science, and engineering. In practice, a huge variety of variants of the TSP occurs in almost every scientific field or application domain: engineering, physics, biology, life sciences, and manufacturing, just to name a few. 

Although the ``general'' TSP is NP-hard, constraints on the tour or the targeted cities can lead to problem variants for which polynomial-time exact solution or approximation algorithms exist. The existence of lower and upper approximation bounds helps to further classify problem variants. Lower and upper bounds characterize the difficulty of a variant and indicate how well a variant can be solved with an efficient, \ie polynomial-time, algorithm. In particular, most upper-bound results, which we review in this paper, devise such an algorithm that guarantees a certain solution quality. It is of high relevance to researchers and practitioners alike to easily determine whether a particular result applies to a variant under study. 

The amount of papers published about the TSP is impressive. Searching for ``traveling salesman'' (with quotes) on Google Scholar returned about 235,000 publications in total.\footnote{Accessed on March 31, 2025.} 19,200 of these have been published since 2020. Searching for ``traveling salesperson'' (with quotes) returns 3,220 publications published since 2020. These publications reveal the almost endless number of variants being studied. It is not only the sheer number of publications, but in particular the many variants in which the TSP is studied that complicates further progress in research and the transfer of theoretical results into practice.

When taking a closer look at some publications, one realizes that is often difficult to precisely identify the variant that it is studied due to the non-uniform terminology that is used when describing a variant. As a consequence, it is hard to find related work because the same problem variant may be named differently, or different problem variants may be named similarly or even identically. The non-uniform representation of variants makes it challenging to understand whether a theoretical or empirical result applies to the variant under consideration as details of the problem definition often remain unclear or have to be identified by carefully reading formal proofs where additional assumptions are described. The current ``muddle'' of variant definitions also hinders the communication between interested parties and the transfer of theoretical results into practice.  Notably, practitioners find it impossible or far too time-consuming to position their problem w.r.t.\ a known problem variant and to benefit from existing research results. It is too cumbersome to judge whether an approximation result is applicable to a specific problem variant and whether or not an algorithm can be reused if a review of the proof behind the result is required.

In this paper, we review approximability results for TSP variants and propose a uniform, systematic and extensible definition scheme named \tico (pronounced \textipa{ti: \textglotstop Es pi: ti:ko}) that we believe is equally well-suited to satisfy the needs of researchers from computer science, economics, and mathematics as well as practitioners.  \tico represents the five parameters, which we use to characterize TSP: the \textbf{T}raveler, the  \textbf{T}argets, the  \textbf{T}our, the  \textbf{C}osts, and the  \textbf{O}bjectives. The definition scheme takes inspiration from similar scheme(s) from the domain of scheduling problems~\cite{conway1967,graham1979,pinedo2012}, which use a tuple of parameters for the characterization of a scheduling variant, where each parameter is characterized by one or several attribute-value fields. Existing work on scheduling classification schemes also shows that slightly different classification schemes have emerged from the early work introduced in~\cite{graham1979} and that it is not always fully clear which scheme is adopted in a scientific publication as the schemes do not have a clearly recognizable name and the definition of attributes and values may slightly differ. The lack of a name also makes it impossible to directly search for publications using a specific scheme. Therefore, we decided to name the proposed definition scheme for TSP variants \tico. We also opted for strict mathematical rigor when defining the tuple of parameters and their associated attributes and values. We also added the year to the  \tico name and speak of "\tico 2025" to support the versioning of the definition scheme and to facilitate a concise reference in publications as the set of attributes and values used to specify a parameter may change over time.

This paper is organized as follows: Section~\ref{section.tsp-intro} introduces the TSP and gives basic definitions as well as an example to illustrate the definitions. Section~\ref{section.related} discusses earlier reviews of approximability results and reviews related work on the definition and taxonomic classification of problem variants for scheduling, traveling salesperson, and vehicle routing problems as well as selected subfields of artificial intelligence research. Section~\ref{section.tico} formally introduces the definition scheme \tico using an EBNF grammar. We distinguish a longhand and a shorthand notation for the definition scheme, the former supporting a detailed, well-structured, and easily readable variant definition and the latter allowing for a very compact precise definition of TSP variants. In Section~\ref{section.attributesvalues}, attributes and values for the fields of the \tico 2025 definition scheme are formally defined. These attributes and values are used to characterize TSP variants in our review of approximability results. In Section~\ref{section.approximability}, we present our review of approximability and inapproximability results for TSP variants.  We use \tico 2025 to define commonly known variants of the TSP that are widely studied, which include the "standard" TSP,  Path TSP, Bottleneck TSP, Maximum Scatter TSP,  Generalized TSP, Clustered TSP, Traveling Purchaser Problem, Profitable Tour Problem, Quota TSP, Prize-Collecting TSP, Orienteering Problem, Time-dependent TSP, TSP with Time Windows, and the Orienteering Problem with Time Windows. We contacted the authors of the referenced publications as far as possible and asked them to review the \tico definition of their respective variant. Many of them got in touch with us, which allows us to present over 60 confirmed definitions that are marked with $\confirmed$.  In Section~\ref{section.vrp}, we briefly discuss how \tico 2025 can be extended to define vehicle routing problem variants involving more than one traveler. Finally, in Section~\ref{section.conclusion}, we summarize the contributions of this paper.  Appendix~\ref{appendixA} provides a cheat sheet for \tico 2025.
In Appendix~\ref{appendixB}, we propose definitions for the TSP variants surveyed in Section~\ref{section.approximability}.

\section{The Traveling Salesman Problem}
\label{section.tsp-intro}

TSP is one of the best-known combinatorial optimization problems. Given a list of cities and distances between them, a traveling salesperson needs to find a shortest possible route to visit the cities. There are many variants of this problem, which makes it hard to formally define a generic version. However, many variants of TSP (if not all) can be formulated over a graph. Let $G=(V,E)$ be a graph with $n$ nodes in $V$ representing the cities and $m$ edges in $E$ representing connections between cities. The edges can either be directed or undirected where the latter implies that the direction of travel does not matter, and the former allows us to model one-way roads and asymmetries, \eg between uphill and downhill travel times. An undirected edge $e$ is an unordered two-element subset of the node set, \ie $e=\{v,w\} \subseteq V$. If it is directed, it is an ordered pair of nodes, \ie $e=(v,w)$, which means that the edge is directed from $v$ to $w$. Generally speaking, TSP are concerned with \emph{walks} in the given graph, which can formally be defined as follows, see~\cite{schrijver2003combinatorial}.

\begin{definition}[Walk]
A walk in a graph $G=(V,E)$ is a sequence
\[
\sol = \left(v_0,e_1,v_1,\ldots,e_k,v_k\right),
\]
where $e_i=\{v_{i-1},v_i\}$ or $e_i=(v_{i-1},v_i)$ for all $i \in [k] := \{1, \ldots, k\}$ in an undirected or directed graph, respectively.
A walk is called \emph{closed} if $v_k=v_0$. A walk is called \emph{simple} if no nodes appear more than once in the sequence except for potentially $v_0=v_k$. A simple closed walk is called \emph{circuit}. A simple walk that is not  a circuit is called \emph{path}.
\end{definition}

A walk in an undirected graph implicitly defines a direction for the edges on the walk. A circuit containing all nodes of the graph is usually referred to as a \textsc{Hamiltonian Circuit} and a path containing all nodes is accordingly called \textsc{Hamiltonian Path}.

There are variants of TSP that distinguish between \emph{visiting} and \emph{traversing} a node, \eg depending on whether a traveler stops in a city or just passes through. To this end, we allow to drop nodes from a walk, \ie more than one edge may appear in the sequence between two nodes provided that consistency\footnote{Should a walk consist solely of undirected edges between the same two nodes, dropping both nodes may lead to an ambiguity about the direction. In this case, we assume that the originating proper walk can be restored from the context.} is maintained. In this light, $v \in \sol$ refers to a node that is visited by $\sol$, \ie explicitly appears in $\sol$. We call a walk proper if it is an alternating sequence of nodes and edges. If a walk $\sol$ is not proper, then we call the inclusion-wise maximal walk with the same set of edges the \emph{originating proper walk} of $\sol$. Let $(v_1,\ldots,v_k)$ be the sequence of visited nodes of some walk. Accordingly, we define $visits_\sol(v) := |\{i: v_i = v\}|$ as the multiplicities in $\sol$ for all $v \in V$. If the walk is clear from the context, we may omit the subscript $\sol$. Moreover, we define the sequence of traversed nodes of $\sol$ as the sequence of visited nodes of the originating proper walk $\mathcal{W}$. Accordingly, we have $traversals_\sol(v) := visits_\mathcal{W}(v)$. If no distinction is made between visited and traversed nodes, then, as the default, the node is traversed.

In what follows, we sometimes refer to parts of a walk. For convenience, we define the following notation.

\begin{definition}[Parts of a Walk]
Let $\sol := (v_0,e_1,\ldots,v_k)$ be a walk. We use the notation
\begin{itemize}
    \item $\sol_V$ for the sequence of visited nodes of a walk,
    \item $V_\sol := \{ v \in V: v \in \sol \}$ for the set of visited nodes of a walk,
    \item $\sol_E$ for the sequence of edges of a walk,
    \item $E_\sol := \{ e \in E: e \in \sol\}$ for the set of traversed edges of a walk,
    \item $\sol_{\leq i} := (v_0,e_1,\ldots,v_i)$ for the prefix of a walk ending at its $i$-th visited node where $i \leq k$,
    \item $\sol_{<i} := \sol_{\leq i} - v_i$ where the last node is removed from the sequence,
    \item $\sol_{i} := \sol_{\leq i} - \sol_{< i}$, which is the $i$-th visited node of the sequence,
    \item $start(\sol)$ is the first traversed node of $\sol$, \ie $start(\sol) = \mathcal{W}_V(0)$ where $\mathcal{W}$ is the originating proper walk of $\sol$,
    \item $end(\sol)$ is the last traversed node of $\sol$, \ie $end(\sol) = \mathcal{W}_V(k)$ where $\mathcal{W}$ is the originating proper walk of $\sol$ and $k = |\mathcal{W}_V|-1$.
\end{itemize}
\end{definition}

The difference between $\sol_V$ and $V_\sol$ is that the former is a sequence with potentially multiple occurrences of some nodes, while the latter is a set of nodes without multiplicities. This distinction is necessary to model the objective for some TSP variants.

A \emph{cost function} $c$ can be any map from the set of walks to the reals. However, it is most common to consider separable functions that are sums over costs of visited nodes and edges of the walk where occurrences are counted with multiplicities, \ie
\[
\cost(\sol) = \underbrace{\sum_{e \in \sol_E} \cost_1(e)}_{=: \cost_2(\sol_E)} + \underbrace{\sum_{v \in \sol_V} \cost_2(v)}_{=: \cost_1(\sol_V)},
\]
where $c_1: E \to \mathbb{R}$ and $\cost_2:V \to \mathbb{R}$. Moreover, there are variants that do not require to visit all nodes, but penalties $p$ occur for nodes that are not visited. For convenience, we use the notation
\[
\compl{p}(\sol_V) := \sum_{v \in V \setminus V_\sol} p(v).
\]
For a directed edge $(u,v) \in E$, we will use the notation $c(u,v)$ to refer to its cost instead of $c((u,v))$. Similarly, we write $c(u,v)$ for an undirected edge $\{u,v\}$ instead of $c(\{u,v\})$, where it is implicitly assumed that $c(u,v)=c(v,u)$. Moreover, $c(\sol)$ always refers to the total costs even if the domain of $c$ is not the set of walks, \ie we consider a suitable lifting of $c$. For some TSP variants, it is necessary to consider parallel edges between the same pair of nodes, \eg when their costs are time-dependent or multiple cost functions are involved where no edge dominates the other in all aspects.

\begin{figure}[htp]
\begin{minipage}[c]{0.5\textwidth}
    \includegraphics[width=\textwidth,clip]{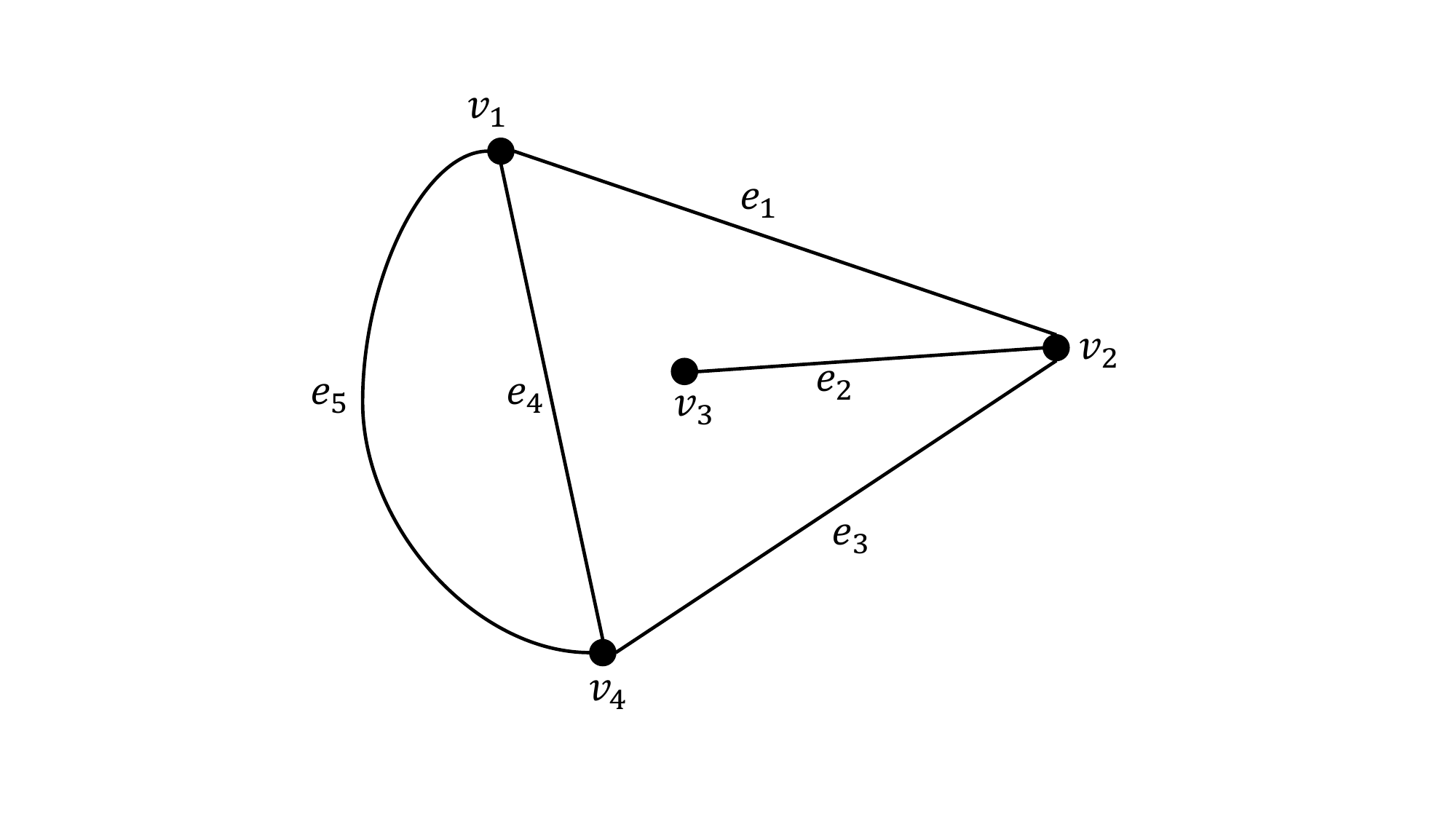}
\end{minipage}
\hspace{1mm}
\begin{minipage}[c]{0.2\textwidth}
    \begin{tabular}{|c|c|} \hline
        edge $e$  &  $c_1(e)$\\ \hline
        $e_1$  & 2 \\
        $e_2$  & 1 \\
        $e_3$  & 2 \\
        $e_4$  & 2 \\
        $e_5$  & 3 \\\hline
    \end{tabular}
\end{minipage}
\hspace{1mm}
\begin{minipage}[c]{0.2\textwidth}   
     \begin{tabular}{|c|c|} \hline
        node $v$  &  $c_2(v)$\\ \hline
        $v_1$  & 1 \\
        $v_2$  & 1 \\
        $v_3$  & 1 \\
        $v_4$  & 1 \\ \hline
    \end{tabular}
\end{minipage}
    \caption{\label{fig:graph}Example of an undirected graph with its edge and node costs.}
\end{figure}

Figure~\ref{fig:graph} shows an example graph of a TSP with $V=\{v_1,v_2,v_3,v_4\}$ and $E= \{e_1,e_2,e_3,e_4,e_5\}$ as well as associated edge and node costs. Let us consider the following three walks in the graph from Figure~\ref{fig:graph}: $\sol_1 = (v_1, e_1, v_2, e_2, v_3, e_2, e_3, v_4, e_4)$, in which node $v_2$ is visited before node $v_3$, the walk $\sol_2 = (v_1, e_1, v_2, e_2, v_3, e_2, v_2, e_3, v_4, e_4)$, in which node $v_2$ is visited twice, and the walk $\sol_3 = (v_1, e_1, v_2, e_2, v_3, e_2, e_3, v_4, e_5)$, which uses an alternative edge $e_5$ to close the circuit. The sequence $\sol_E = (e_1, e_2, e_2, e_3, e_4)$ describes the edges of the first two walks. In this sequence, visited vertices remain implicit and the sequence leaves open whether the traveler visits node $v_2$ before or after node $v_3$ or even twice.  The costs of these walks in our example are $\cost(\sol_E)=8$,  $\cost(\sol_1)=12$, $\cost(\sol_2)=13$, and $\cost(\sol_3)=13$. In a TSP variant where we do not care about node costs and only consider the length of the tour, $\sol_E$ is sufficient as a representation of a solution.

\section{Related Work}	
\label{section.related}

Until today, there is no comprehensive review of polynomial-time approximability and inapproximality results available. Papers that publish new theoretical results usually limit their related work section to earlier work with results for the TSP variant studied in the paper or on publications of algorithmic techniques that are used or improved to achieve a new result. Excellent examples are ~\cite{vygen2012}, which focuses on the asymmetric, symmetric, and graphic variants of the "standard" TSP, and ~\cite{traub2020approximation}, which focuses on asymmetric and symmetric variants of the "standard" TSP and the path TSP.
A recent and comprehensive survey on the Generalized TSP was published in~\cite{pop2023comprehensive}, which provides an overview on exact and approximate algorithms, transformation methods, and heuristic and metaheuristic algorithms. A survey on heuristic approximations was published in \cite{kongkaew2014survey}. Older surveys are~\cite{BDDVW1998}, which resents an overview of polyniomal-time solvable TSP variants known until 1998, and~\cite{laporte1992traveling, bern1997approximation}, which review selected exact and approximate algorithms for the "standard" TSP. A very good overview on the TSP and a bibliography of historic TSP publications is provided by William Cook.\footnote{\url{https://www.math.uwaterloo.ca/tsp/index.html}}

With the growing number of publications on the TSP in the 1990s, an increasing desire to classify problem variants and to clarify the relationships between  published results can be observed. A paper published in 1991 by Langevin et al.~\cite{langevin1990} reviews and compares different formulations of the TSP with the goal of establishing correspondences between the different published results. Another analysis and review of publications on time-dependent TSP variants was published in 1995 by Gouveia and Vo\ss~\cite{gouveia1995classification}. A bibliography of 500 publications on TSP, vehicle routing, chinese postman, and rural postman problems was published in 1995 by Laporte and Osman~\cite{laporte1995}, which was motivated by ``the abundant and somewhat disorganized literature'' on these problems. This paper used a ``simple and relatively broad classification scheme''  specifying the properties of the graph, the objective, and a set of constraints. However, none of the attributes or their values were formally defined because the focus of this classification scheme was more on providing a taxonomy to organize and structure the comprehensive list of references collected in the paper. 

A successful attempt to classify TSP variants was published in 1991 by Reinelt with the Traveling Salesperson Problems Library TSPLIB~\cite{reinelt1991}. TSPLIB had the goal of providing the community with challenge problems and solutions. It was, to our knowledge, the first attempt that introduced specific attribute/value pairs to define TSP variants. For example, the keyword \texttt{TYPE} with possible values such as \texttt{TSP}, \texttt{ATSP}, or \texttt{CVRP} identifies symmetric TSP, asymmetric TSP, or vehicle routing problems. The attribute  \texttt{CAPACITY} is used to denote the capacity of vehicles in a vehicle routing problem,  \texttt{GRAPH\_TYPE} specifies whether the graph is complete or sparse, and \texttt{EDGE\_TYPE} specifies whether edges are directed or undirected. Costs are defined in the \texttt{EDGE\_WEIGHT\_TYPE} attribute and can comprise different distance metrics such as euclidean, Manhattan, or geographical distances. Some of Reinelt's keywords and values also occur in our definition scheme. 

In the decades following the work on TSPLIB, many new  TSP variants were studied; however, less interest can be observed in the community to adopt a uniform scheme for defining these variants. The book by Applegate et al.~\cite{applegate2011} from 2011 touches upon ``the many flavors of the TSP'' in its introductory chapter but only consider the ``standard'' TSP that is equivalent to computing a Hamiltonian circuit. In~\cite{deineko2014}, a complete classification for restricted polynomially-time solvable variants of the TSP with so-called four-point conditions is presented  to establish a precise border between polynomially-time solvable and NP-hard variants of this restricted problem. 

For the TSP with several travelers, also known as the multiple TSP or MTSP, and commonly denoted as the vehicle routing problem (VRP), several proposals for a classification scheme or taxonomy have been made over the last decades. They are motivated by the high practical relevance of the problem and the huge number of variants in which a VRP can occur in practice. Interestingly, work on systematizing VRP variants is more focused on defining taxonomies rather than on formally defining the properties that characterize a variant. Researchers implicitly seem to take it for granted that there is a uniform and precise understanding of a problem variant and, therefore, naming the problem variant is sufficient for its precise identification. 

The first VRP taxonomy dates back to Bodin~\cite{bodin1975} in 1975, who observed that ``this problem has a reasonably natural taxonomic structure'' and that ``many vehicle routing and scheduling problems can be classified by their underlying properties''.  Bodin introduced three characteristics of a VRP: First, the network classification that specifies whether it is a node routing, branch routing, or general routing problem. Second, the number of vehicles, and third, the type of the algorithm or solution method, which is used to solve a problem, \eg exact or heuristic methods. In \cite{bodingolden1981}, the attributes to characterize a VRP variant were refined into 13 different categories and 7 different types of solution methods were distinguished. All later proposals for the classification of VRPs are variations of this early work.

In 1990, Desrochers et al. \cite{desrochers1990} introduced a TSP classification scheme with four fields: \textit{addresses}, \textit{vehicles}, \textit{problem characteristics}, and \textit{objectives}. A wide variety of attributes and possible values is introduced for each field, but the paper does not provide a formal definition  for the attributes. To specify the scheme, a context-free grammar was used and several examples of well-known VRP and TSP were classified. For example, the TSP with time windows is denoted by $1, tw_j \mid 1 \mid \Delta \mid T$. The proposal nevertheless was not successful, perhaps due to the lack of formal definitions for the attributes and values. Several different  taxonomies were published over time, many of them with the goal of systematically reviewing the abundant literature on TSP/VRP variants; for example, comprehensive reviews on transportation network design and routing problems can be found in~\cite{current1986,current1993}.  Dynamic VRP variants where ``part of the input required to solve it (that is, which nodes actually request service) is revealed to the dispatcher concurrently with the determination of the route'' are reviewed in~\cite{psaraftis1995}. Variants of the location-routing problem LRP that combines tour planning and facility location planning within the same problem are reviewed in~\cite{min1998}. An overview on variants of the dynamic resource transformation problem can be found in~\cite{powell2001}. A critical review of these classification proposals was published in~\cite{eksioglu2009} in 2009, which proposes yet another taxonomy to classify so-called generalized routing problems. This taxonomy is meant to be extensible and provides the following 7 classes:  (1) shortest path problem, (2) Chinese postman problem, (3) rural postman problem, (4) dial-a-ride service route problem, (5) arc routing problem, (6) TSP, and (7) VRP. None of these problem classes is formally and precisely defined. 

The literature survey in \cite{pillac2013} from 2013 reviews VRP application areas and algorithm types based on the taxonomy from~\cite{psaraftis1995} and simplifies it into a 4-quadrants model using information quality (stochastic/deterministic) and information evolution (static/dynamic) as its two axes. A  literature review of green vehicle routing problems can be found in~\cite{lin2014}, which introduces a large set of names for various problem variants, but again without providing precise formal definitions. Further taxonomic variations are introduced in these VRP surveys ~\cite{montoya2015,lahyani2015,psaraftis2016,braekers2016,Cheikhrouhou2021} published between 2015 and 2021. A more formally defined classification scheme of VRPs can be found in~\cite{irnich2014} from 2014. Formal definitions for various VRP variants are given and problems are classified based on the (road) network structure, the type of transportation requests, constraints that affect each route individually, the fleet composition and location, inter-route constraints, and optimization objectives. Formal definitions of attributes and values to characterize the problems along these properties are not given. Instead, the paper informally reviews possible definitions from related work.

The literature offers all these proposals, none of which seem to have convinced researchers and practitioners. As evidence, take for example the fact that none of the available proposals is uniformly adopted by all papers within the research paper collections published between 2008 and 2020~\cite{golden2008,Toth2014,derbel2020}. We see several reasons for this ongoing quest for an easy-to-use and extensible approach to precisely characterize TSP/VRPs: First, a taxonomy introduces a predefined hierarchical structure to align or separate problem classes, which is too rigid to cover the manifold variants under study. Secondly, most of the proposals introduce problem attributes without providing precise mathematical definitions of these attributes and their possible values. This has the consequence that the classification of a problem instance often depends on the informal problem description of a problem class, which might give room for interpretation. Thirdly, most VRP taxonomies serve as a basis to organize a survey of the scientific literature and are less aimed at providing a formalism that would support researchers and practitioners in the precise definition of their problem variant. Last but not least, we believe that the definition of the problem and possible solution methods/algorithms should be clearly separated. However, as most VRP taxonomies are used to review the scientific literature, it is not surprising that adopted solution methods are an essential part of most taxonomies.

The work on \tico is inspired by classification schemes used to characterize scheduling problems. A first such scheme was introduced by Conway, Maxwell, and Miller in their seminal book from 1967~\cite{conway1967}. It uses four information categories to describe a scheduling problem, where each category introduces different attributes and possible values and is denoted with an upper-case letter: \textit{A} - jobs and operations to be processed, \textit{B} - number and types of machines available for processing, \textit{C} - the variant of the scheduling problem such as, for example, flow-shop or randomly routed job-shop, \textit{D} - evaluation criteria (objective) for the evaluation of a schedule such as, for example, the average or maximum completion time, flow time, lateness or tardiness. A notation was defined that separates each category by a slash and lists the attribute values. As examples for this notation, the problem $n/2/F/F_{max}$ asks to minimize the maximum flow-time, which is the time the last job is finished, for $n$ jobs in a 2-machine flow shop, whereas the problem $n/m/G/F_{max}$ asks to schedule $n$ jobs in a job shop with $m$ machines using the same objective.

It took some time for this classification scheme to gain influence. A survey paper~\cite{BakshiArora1969} from 1969 and the textbook by Baker from 1974~\cite{baker1974} do not make use of it although both give a comprehensive overview on classes of scheduling problems. The collection edited by Coffman~\cite{coffman1976} in 1976 also provides a detailed review of the state of the art in scheduling algorithms in the 1970s but uses its own scheduling model with the elements \textit{resources}, \textit{task systems}, \textit{sequencing constraints} (focusing on preemption/non-preemption and task priorities), and \textit{performance measures}. A \textit{task system} is specified by a set of tasks, precedence constraints, execution times, resource requirements, and weights being interpreted as deferral costs. The model is described using a specific notation and was used by all contributing authors of this book.

It was not until the paper by Graham et al.\ in 1979~\cite{graham1979} that a 3-field classification scheme $\alpha \mid \beta \mid \gamma$ was introduced, but it does not explicitly refer to the work by Conway~\cite{conway1967}. In this scheme, $\alpha$ characterizes the machine environment, $\beta$ specifies the job characteristics, and $\gamma$ refers to the objective. Similarly to~\cite{conway1967}, a shorthand notation was introduced that separates fields by vertical lines and only lists attribute values between the lines. In this notation, the example $1 \mid prec \mid L_{max}$ specifies the problem to minimize the maximum lateness of jobs with precedence constraints on a single machine. We build on the idea of clearly separated fields denoted by Greek letters and also make use and extend the idea of the shorthand notation. 

Comparing both~\cite{graham1979} and~\cite{conway1967}, one notices that the $\alpha$-field combines information from the B- and C-field and that the $\gamma$-field corresponds to the D-field. The  $\beta$-field is a new field and is refined further to specify job characteristics such as preemption  or required resources. The data that defines a specific problem instance is separated as so-called ``job data'' and is not part of the classification scheme defined in~\cite{graham1979}, whereas in~\cite{conway1967} it is part of the A-field in the classification scheme. The book chapter by Lawler et al.~\cite{lawler1993} in 1993 makes extensive use of the classification scheme by Graham et al~\cite{graham1979}. The textbook by Brucker~\cite{brucker1995} from 1995 uses, refines, and slightly modifies this classification scheme further. For example, additional attributes are introduced for the $\alpha$-field to characterize more complex machine environments involving multi-processor tasks. Both publications~\cite{lawler1993,brucker1995} also make extensive use of the shorthand notation introduced in~\cite{graham1979} to specify a wide range of problem variants and to provide an overview of known complexity-theoretic results and algorithms for these variants. 
 
Today, the classification  scheme $\alpha \mid \beta \mid \gamma$  is widely established to characterize scheduling problems and is used, for example, in~\cite{dawande2005sequencing, pan2013comprehensive, pfund2004survey, ruiz2005comprehensive, sayadi2010discrete} to name a few  publications with an overview character. In the influential textbook by Pinedo~\cite{pinedo2016}, the $\alpha$-field contains a single entry with one out of 9 possible values to characterize the machine environment. The $\beta$-field is more flexibly used to specify processing characteristics and constraints and can contain several entries that are separated by ``,''. Pinedo lists an initial set of 12 possible entries emphasizing that ``any other entry that may appear in the $\beta$-field is self explanatory''~\cite[p. 17]{pinedo2016}. The $\gamma$-field describes the single (and to be minimized) objective. In the handbook on scheduling by B{\l}a{\.z}ewicz et al.~\cite{blazewicz2019} from 2019, the $\alpha \mid \beta \mid \gamma$ classification scheme is used with the $\alpha$-field being refined into two subfields and the $\gamma$-field being refined into 8 subfields.

Other fields tackling optimization problems have also developed classification schemes: a taxonomy to classify black-box and simulation-based optimization problems based on constraint classes is proposed in~\cite{digabel2015}, multi-disciplinary design optimization architectures are surveyed and classified in~\cite{martins2013}, and a classification of formulations for the optimization of distributed systems can be found in~\cite{tosserams2009}. A comprehensive overview on graph problems and the P/NP boundary can be found in~\cite{isgci}.

Researchers in Artificial Intelligence have also developed formally defined definition schemes and taxonomies to characterize problem variants. A well-known example is the work on ontological representations using so-called description logics (also known as concept languages or terminological logics), which are fragments of first-order predicate logic. A large family of language variants is defined with the goal to determine decidable fragments of first-order predicate logic and to study the complexity of reasoning algorithms deciding for example subsumption in these fragments,  see~\cite{baader2007,baader2017} for an overview. The theoretical work on the decidability and complexity of various description-logic variants has provided the foundation for the semantic web~\cite{Berners-Lee2001} and the Ontology Web Language OWL~\cite{owl2012}. In AI planning, the Planning Domain Definition Language PDDL was introduced in 1998~\cite{mcdermott2000} and extended in 2003~\cite{foxlong2003,mcdermott2003} to achieve a standardized representation of planning problems as the foundation to hold a regular competition among different planning algorithms.\footnote{See https://www.icaps-conference.org/competitions/} PDDL fragments of different expressivity were identified and defined the tracks within this competition. The definition of the PDDL language was preceded by research results on the complexity of planning problems in different planning formalisms, \eg~\cite{bylander1991, backstrom1995}. Research results aiming at understanding the border between tractable and intractable (NP-hard) subclasses of problems were published for temporal reasoning using Allen's Interval Algebra~\cite{nebel95} and for spatial reasoning using the Region Connection Calculus~\cite{renz99}.

\section{The TSP Definition Scheme \tico}
\label{section.tico}

The TSP definition scheme \tico is defined by the EBNF grammars\footnote{Following the EBNF style introduced in~\cite{pattis1994}, we use ( $\mid$ ) to list choices and $\{ \}$ to denote repetitions of 0 or more occurrences. Non-terminal symbols are written in italics and terminal symbols are written within ``''. Strings in normal font such as `mathematical expression', `string', or `set' have the standard meaning from computer science.} in Figures~\ref{fig:ebnf-long} and~\ref{fig:ebnf-short}, which introduce a longhand and a shorthand notation for the definition of a TSP variant using five different fields: The $\alpha$-field characterizes the travelers, who are usually salespeople in TSP, but can also be vehicles, robots, or any other entity that can move and is not stationary. The $\beta$-field describes how often (each or a particular) node in the graph needs to be visited by a traveler. The $\gamma$-field describes the tours or constraints on possible tours a traveler may take. The $\delta$-field describes the cost functions existing in the problem. Finally, the $\epsilon$-field describes the objective(s) in the form of optimization criteria or bounds on the costs. As already mentioned, these five fields give the name to the definition scheme \tico: the \textbf{T}raveler, the  \textbf{T}argets, the  \textbf{T}our, the  \textbf{C}osts, and the  \textbf{O}bjectives.

\begin{figure}[H]
\fbox{
\parbox{0.8\linewidth}{
\begin{tabbing}
\tico \hspace{10mm} \=  $\Leftarrow$ \hspace{5mm} \=``$\langle \,$'' $\alpha$-\textit{field} \hspace{1mm} $\beta$-\textit{field}  
\hspace{1mm} $\gamma$-\textit{field} \hspace{1mm} $\delta$-\textit{field} \hspace{1mm} $\epsilon$-\textit{field} $\, ``\rangle$''\\
$\alpha$-\textit{field} \>  $\Leftarrow$ \>   (``$\alpha$'' $\mid$ ``traveler'')  ``:''   $\{$ \textit{attribute} ``;''  $\}$ \\
$\beta$-\textit{field} \>  $\Leftarrow$ \>   (``$\beta$'' $\mid$ ``target'')  ``:''   $\{$ \textit{attribute} ``;'' $\}$ \\
$\gamma$-\textit{field} \>  $\Leftarrow$ \>   (``$\gamma$'' $\mid$ ``tour'')  ``:''   $\{$ \textit{attribute} ``;''  $\}$ \\
$\delta$-\textit{field} \>  $\Leftarrow$ \>   (``$\delta$'' $\mid$ ``costs'')  ``:''   $\{$ \textit{cost function} ``;'' $\}$ \\
$\epsilon$-\textit{field} \>  $\Leftarrow$ \>   (``$\epsilon$'' $\mid$ ``objective'')  ``:''   
 $\{$\textit{objective} ``;''$\, \}$ \\
\textit{attribute} \> $\Leftarrow$\> \textit{name}  $\,$ \textit{value}\\
\textit{name}\> $\Leftarrow$ \> string\\
\textit{value}\> $\Leftarrow$ \> \textit{relation} $\,$ mathematical expression \\
\textit{relation} \> $\Leftarrow$ \> ``='' $\mid$ ``$\leq$'' $\mid$ ``$<$'' $\mid$ ``$\geq$'' $\mid$ ``$>$'' $\mid$ ``$\in$''\\
\textit{cost function} \>  $\Leftarrow$ \> \textit{name} ``:''  \textit{domain} ``$\mapsto$'' \textit{range}  $\,   \{$``,'' \textit{attribute}$\}$  \\
\textit{domain} \> $\Leftarrow$ \> set\\
\textit{range} \> $\Leftarrow$ \> set\\
\textit{objective} \> $\Leftarrow$ \> mathematical expression
\end{tabbing}
	}
}
	\caption{EBNF for the \tico longhand notation.}
	\label{fig:ebnf-long}
\end{figure}

In the \tico longhand notation, a pair of angle brackets encloses the five fields. The description of each field begins with either a Greek letter or a keyword denoting the field's name followed by a colon. In the $\alpha$, $\beta$, and $\gamma$ field, a sequence of attribute-value pairs separated by semicolons follows. An attribute is described by a name, a relation, and a value. Names can be arbitrary strings, and values can be arbitrary mathematical expressions, as \tico is intended to be extensible and thus does not impose a specific notation for attribute-value pairs. Accordingly, also the list of relations to be used between names and values shown in Figure~\ref{fig:ebnf-long} is not meant to be final. Attribute-value pairs should be meaningful mathematical expressions; for example, the $\in$-relation should only be used with a set as a possible value. 

In the $\delta$-field, the colon is followed by one or more definitions of cost functions. A cost function is defined by a name, followed by a colon after which the domain and range of the cost function are defined. An optional sequence of attribute definitions may follow, which is separated by a comma from the definition of the cost function(s). Domain and range specifications can be arbitrary sets and no specific notation is imposed except the $\mapsto$ arrow between them.

In the $\epsilon$-field, the colon is followed by a sequence of objective functions separated by semicolons. An objective function can be any  mathematically meaningful combination of cost functions from the $\delta$-field and, hence, is only defined as an arbitrary string by the EBNF. Usually, the objective may be either to minimize, maximize or bound (above or below) the objective function(s) or any combination thereof. 

The example below illustrates the longhand notation by defining a variant of the ``standard'' TSP.\footnote{See Section~\ref{sec.standardtsp} for a more detailed discussion of the ``standard'' TSP. Attributes and values will be formally defined in Section~\ref{section.attributesvalues}. We give here a brief informal explanation of each attribute as the example is intended to illustrate the EBNF notation.} In this ``standard'' TSP variant, a single traveler (count = 1) visits each of the $n$ nodes of the graph exactly once (visits = 1). The traveler may start at any node (start = False) and has to return to the starting node at the end of the tour (circuit = True, end = False). An undirected edge (undirected = True) exists between any two nodes (complete = True). A cost function $c$ is defined with the set of edges $E$ as its domain and the range being in the non-negative rational numbers.  The objective is to minimize the tour costs $\min c(\sol)$. 

\begin{tabbing}
\hspace{10mm}$\langle$ \= $\alpha:$ \= count $=1$; \\
\> $\beta:$ \> traversals $= 1$; \\
\> $\gamma:$ \> start = False;\\
            \> \> end = False;\\
            \> \> circuit = True;\\
            \> \> graphtype = complete;\\
             \> \>edgetype = undirected;\\
\> $\delta:$ \> c: $E \mapsto \mathbb{R}_{\geq 0}$;\\
\> $\epsilon:$ \>  $\min c(\sol)$; $\rangle$
\end{tabbing}

The EBNF for the shorthand notation of \tico in Figure~\ref{fig:ebnf-short} introduces a very compact notation that omits the names of fields and that separates fields by vertical lines. For this purpose, the EBNF in  Figure~\ref{fig:ebnf-short} redefines attribute definitions, making the name, relation, and value optional. The intended usage is as follows: As long as attribute values are uniquely defined across all attributes and the same value is not used by different attributes, only the value is listed in a field. If the value is non-unique or there is more than one attribute defined for a field sharing the same range of values, the attribute-value pair must be listed completely to achieve a precise definition. For Boolean attributes with values True and False, the attribute name is listed if the attribute has value True. The absence of a Boolean attribute name indicates that the attribute has value False. The definition of the $\delta$ and $\epsilon$ fields remains unchanged except that the name of the fields are omitted. In this shorthand notation, there is no last semicolon to end each field as vertical lines serve as separators between fields.

\begin{figure}[H]
\fbox{
\parbox{0.8\linewidth}{
\begin{tabbing}
\tico \hspace{3mm} \= $\Leftarrow$ \hspace{1mm}  \= ``$\langle$'' \hspace{1mm}
$\alpha$-\textit{field} \hspace{1mm} ``$\mid$'' \hspace{1mm} 
$\beta$-\textit{field}  \hspace{1mm} ``$\mid$'' \hspace{1mm}
$\gamma$-\textit{field} \hspace{1mm} ``$\mid$'' \hspace{1mm} 
$\delta$-\textit{field} \hspace{1mm} ``$\mid$'' \hspace{1mm} 
$\epsilon$-\textit{field} \hspace{1mm} ``$\rangle$''\\
$\alpha$-\textit{field} \> $\Leftarrow$  \> $\{$\textit{attribute} ``;''$\}$ \textit{attribute} \\
$\beta$-\textit{field} \> $\Leftarrow$  \>  $\{$\textit{attribute} ``;''$\}$ \textit{attribute}  \\
$\gamma$-\textit{field} \> $\Leftarrow$  \> $\{$\textit{attribute} ``;''$\}$ \textit{attribute} \\
$\delta$-\textit{field} \>  $\Leftarrow$ \>  $\{$\textit{cost function} ``;'' $\}$ \textit{cost function}\\
$\epsilon$-\textit{field} \>  $\Leftarrow$ \> $\{$\textit{objective} ``;'' $\}$ \textit{objective}\\
\textit{attribute} \> $\Leftarrow$\> (\textit{name} $[$\textit{value}]) $\mid$ ($[$\textit{relation}] mathematical expression)\\
\textit{cost function} \>  $\Leftarrow$ \> \textit{name} ``:''  \textit{domain} ``$\mapsto$'' \textit{range}  $\,   \{$``,'' \textit{attribute}$\}$  
\end{tabbing}
	}
}
	\caption{EBNF for the \tico shorthand notation. Rules for non-terminal symbols not defined here can be found in Figure~\ref{fig:ebnf-long}.}
	\label{fig:ebnf-short}
\end{figure}

The TSP variant, which we defined above in the longhand notation, now simply reads 
\[
\langle \, =1 \, | \, =1 \, | \, \mbox{circuit; complete; undirected} \, | \, c : E \mapsto \mathbb{R}_{\geq 0} \, | \,  \min c(\sol)\,  \rangle
\]
in the \tico shorthand notation. Only listing the relation and numeric value $=1$ without the attribute name in the $\alpha$ and $\beta$-field yields a unique definition of this TSP variant based on the initial list of attributes that we propose for these fields in Section~\ref{section.attributesvalues}. All attributes in the $\gamma$-field are Boolean attributes and therefore only the names of these attributes that all have the value True are listed.

In Section~\ref{section.approximability}, manifold examples of the shorthand notation will be shown. Some of these examples impose additional constraints on a problem variant that are for example exploited in the formal proof of an approximability result. These constraints are often specific to a certain theoretical result. In this case, the \tico shorthand notation ends with  $\rangle^\oplus$ and the constraints are stated separately as text or formal definitions. If several \tico definitions in a publication require extensions, a running number such as $\oplus 1$ can be added. In Section~\ref{section.vrp}, we discuss how \tico can be extended to deal with MTSP and VRPs by introducing additional attributes for the $\alpha$-field. An extension of \tico may invalidate the uniqueness of earlier attribute definitions. For example, it seems natural for a VRP to describe the number and capacities of moving vehicles in the  $\alpha$-field, which would make a definition of just $=1$ in this field non-unique. We therefore use versioning and explicitly state the year of publication of a specific \tico version.

\section{Formal Definition of Attribute-Value Pairs}
\label{section.attributesvalues}	

In this section, we proceed with the formal definition of an initial set of attributes and values r the $\alpha$, $\beta$, $\gamma$, and $\delta$-field that we propose to start with in \tico 2025. As the definition scheme is extensible, we understand these definitions as an initial core set that can be easily extended by researchers, with new values or attributes as needed. We also develop the notation for the $\epsilon$-field further. 
 
\subsection{The Travelers Field \texorpdfstring{$\alpha$}{α}}

The $\alpha$-field describes the number of salespeople traveling simultaneously, or equivalently the number of visiting sequences that are to be found. In the case of multiple salespeople, it may also describe their characteristics, such as capacities. For the ``standard'' TSP, there is just a single salesperson who has to visit given cities. The case of several salespeople leads to the vehicle routing problem, for which we briefly discuss possible extensions of the $\alpha$-field in Section~\ref{section.vrp}.

\medskip

\noindent \textbf{\textit{count}: int | expression}\\
The \emph{count} attribute describes the number of salespeople considered in the problem; for example, the \emph{count} entry of the ``standard'' TSP is $=1$ as exactly one salesperson exists in the problem \cite{dantzig1954solution}. The classical vehicle routing problem \cite{bellmore1974transformation,rao1980note}, on the other hand, has \emph{count} $=k$ for some constant $k$.
\begin{itemize}[align=left,labelwidth=\widthof{$=\Theta(1)$},leftmargin=\labelwidth + \labelsep]
	\item[$=1$] A solution consists of a single finite sequence.
	\item[$=\Theta(1)$] A solution consists of $k$ finite sequences, where $k=\Theta(1)$ is a fixed parameter that is not part of the input.
	\item[$=k$] A solution consists of $k$ finite sequences, where $k\geq 2$ is part of the input.
	\item[$\geq 1$] A solution consists of at least one finite sequence.
\end{itemize}

\subsection{The Targets Field \texorpdfstring{$\beta$}{β}-Field}

The $\beta$-field describes which cities and how often each city has to be visited and describes the targets of the tour. In the ``standard'' TSP, each city has to be visited exactly once or at least once, but other variants exist, which allow for certain cities not to be visited or ask for some cities to be visited multiple times. A \emph{solution} of a TSP is a walk (see Section~\ref{section.tsp-intro}) and the $\beta$-field describes properties a solution must satisfy. 

\medskip

\noindent \textbf{\emph{traversals}: int | expression}\\
The \emph{traversals} attribute describes how often each city is to be traversed in a solution sequence. In the ``standard'' TSP~\cite{miller1960integer}, in which each city must be traversed exactly once, the attribute takes value $=1$. If cities may be traversed multiple times, but must be traversed at least once, the value of the \emph{traversals} attribute is $\geq 1$, which is the case for example in the TSP with multiple visits (TMSP)~\cite{gutin2006traveling}. Furthermore, if each city should be traversed at most once and is allowed to not be traversed at all, the \emph{traversals} attribute takes value $\leq 1$, cf.~\cite{feillet2005traveling}.

\begin{itemize}[align=left,labelwidth=\widthof{$\geq d(v)$},leftmargin=\labelwidth + \labelsep]%
	\item[$=1$] For any solution $\sol$ and all $v \in V$, it must hold that $traversals_\sol(v) = 1$.
	\item[$=d$] For any solution $\sol$ and all $v \in V$ it must hold that $traversals_\sol(v) = d$, where $d\geq 2$.
	\item[$=d(v)$] For any solution $\sol$ and all $v \in V$, it must hold that $traversals_\sol(v) = d(v)$.
	\item[$\geq 1$] For any solution $\sol$ and all $v \in V$, it must hold that $traversals_\sol(v) \geq 1$.
	\item[$\mathit{\geq 0}$] For any solution $\sol$ and all $v \in V$, it must hold that $traversals_\sol(v) \geq 0$.
	\item[$\geq d(v)$] For any solution $\sol$ and all $v \in V$, it must hold that $traversals_\sol(v) \geq d(v)$.
	\item[$\leq 1$] For any solution $\sol$ and all $v \in V$, it must hold that $traversals_\sol(v) \leq 1$.
\end{itemize}
If one wants to further specify the value of an attribute, it is possible to extend the string value. As an example, suppose one wishes to express that for a given subset $D$ of $V$ any solution $\sol$ must traverse each node in $D$ at least once with no restrictions on the number of traversals of other nodes. To express this, the value $\geq d(v)$ may be extended to $\geq d(v)\in\{0,1\}$.

\medskip

\noindent \textbf{\emph{visits}: string}\\
The \emph{visits} attribute describes how often each node is to be visited, \ie how often each node occurs in a solution. Recall from Section~\ref{section.tsp-intro} that a node may be traversed without being visited. If every node is visited on each traversal, it takes value \emph{always}. If each traversed node is only visited once, the attribute takes value \emph{once}.  If the traveler receives a bonus for each visit of a node, but also consumes time when visiting a node, the \emph{visits} attribute may have value $\geq once$ to say each node must be visited at least once, but may also be visited multiple times, or it may have value $\leq once$ to say that some nodes must not be visited.
\begin{itemize}[align=left,labelwidth=\widthof{= \emph{always}},leftmargin=\labelwidth + \labelsep]%
    \item[= \emph{always}] For any solution $\mathcal{S}$ and all $v \in V$, it must hold that $visits_\sol(v) = traversals_\sol(v)$.
    \item[= \emph{once}] For any solution $\sol$ and all $v \in V$ such that $traversals_\sol(v)>0$, it must hold that $visits_\sol(v) = 1$.
    \item[$\geq$ \emph{once}] For any solution $\sol$ and all $v \in V$ such that $traversals_\sol(v)>0$, it must hold that $visits_\sol(v) \geq 1$.
    \item[$\leq$ \emph{once}] For any solution $\sol$ and some $v \in V$ such that $traversals_\sol(v)>0$, it can hold that $visits_\sol(v) \leq 1$.
\end{itemize}
If no distinction is made between visited and traversed nodes, then the visits attribute does not appear in the \tico variant definition. In the short-hand notation, we can leave out the $=$ sign if there is no ambiguity, \ie we write \emph{always} instead of  \emph{=always}.

\medskip
\noindent 
\textbf{\emph{group}: string}\\
The nodes in the graph can be grouped into sets $\{P_1,\dots,P_k\}$. The \emph{group} attribute describes the properties of these groups/sets and how many nodes in each group must be visited. Groups can be disjoint or overlapping and exactly one or at least one node from each group must be visited, which is described by the parameter = once or $\geq$ once in  parentheses behind the property value.
\begin{itemize}[align=left,labelwidth=\widthof{= \textit{partition} $\mbox{($\geq$ once)}$},leftmargin=\labelwidth + \labelsep]
    \item[= \textit{partition} $\mbox{(once)}$] For all $v\in V$, there exists a unique $i\in \{1,\dots,k\}$ such that $v\in P_i$, and for any solution $\sol$ and all $j\in\{1,\dots,k\}$, there exists a unique $u\in V_\sol$ such that $u\in P_j$.
    \item[= \textit{partition} $\mbox{($\geq$ once)}$] For all $v\in V$, there exists a unique $i\in \{1,\dots,k\}$ such that $v\in P_i$, and for any solution $\sol$ and for all $j\in\{1,\dots,k\}$, there exists at least one $u\in V_\sol$ such that $u\in P_j$.
    \item[= \emph{cover} $\mbox{(once)}$] For all $v\in V$, there exists at least one $i\in \{1,\dots,k\}$ such that $v\in P_i$, and for any solution $\sol$ and all $j\in\{1,\dots,k\}$, there exists a unique $u\in V_\sol$ such that $u\in P_j$.
    \item[= \emph{cover} $\mbox{($\geq$ once)}$] For all $v\in V$, there exists at least one $i\in \{1,\dots,k\}$ such that $v\in P_i$, and for any solution $\sol$ and all $j\in\{1,\dots,k\}$, there exists at least one $u\in V_\sol$ such that $u\in P_j$.
\end{itemize}
The set of parameters may also be extended to include further information, for example specify the number of nodes that should be traversed in each cluster or how many clusters exist in the problem. Further, the number of clusters may also be described by replacing the \emph{partition} value by \emph{bipartition} (if there are exactly two clusters) or $k$-partition (if there are exactly $k$ clusters). Examples of such value specifications can be found in Section~\ref{sec.approx.generalized}.

\medskip
\noindent \textbf{\emph{covering}: string}\\
The \emph{covering} attribute describes the subset of nodes that is to be covered by a solution walk, where a node $v$ is covered by a solution $\sol$ if it is either on the tour, \ie $v\in V_\sol$, or within a predetermined cost bound to an arbitrary node on the tour. The values include two parameters. The first parameter $c$ is the cost function used to measure the shortest-path distance from a node to the solution walk, where $c$ must be a cost function or expression over cost functions defined in the $\delta$-field. The second parameter gives the bound $b$ on how far nodes are allowed to be from the solution walk.
\begin{itemize}[align=left,labelwidth=\widthof{= \emph{subset} $(c, \leq b)$},leftmargin=\labelwidth + \labelsep]
    \item[= \emph{all} $(c, \leq b)$] For any solution $\sol$ and for all $v\in V$ there exists $u\in V_\sol$ such that, $u=v$ or $dist_c(u,v)\leq b$.
    \item[= \emph{subset} $(c, \leq b)$] For any solution $\sol$ and for all $v\in D\subset V$ there exists $u\in V_\sol$ such that, $u=v$ or $dist_c(u,v)\leq b$.
\end{itemize}

\subsection{The Tours Field \texorpdfstring{$\gamma$}{γ}}

The $\gamma$-field characterizes the graph of the problem and defines possible constraints on the tour such as a fixed start or end node or precedence constraints. In many variants of the TSP, the underlying graph is complete, \ie every node can be reached from every other node, the salesperson needs to return to its origin node at the end and neither start nor end node are fixed. 

\medskip
\noindent \textbf{\emph{start}: bool}\\
The \emph{start} attribute defines whether the starting node of the tour is fixed as for example in~\cite{an2015improving} or not.

\begin{itemize}[align=left,labelwidth=\widthof{= \emph{False}},leftmargin=\labelwidth + \labelsep]
    \item[= \emph{True}] Given start node $s$, every solution $\sol$ must satisfy $start(\sol)=s$.
    \item[= \emph{False}] For any solution $\sol$, $start(\sol)$ can be any node in $V$.
\end{itemize}

\medskip
\noindent \textbf{\emph{end}: bool}\\
The \emph{end} attribute defines  whether the end node of the tour is fixed, as, for example, in~\cite{an2015improving}, or not.
\begin{itemize}[align=left,labelwidth=\widthof{= \emph{False}},leftmargin=\labelwidth + \labelsep]
    \item[= \emph{True}] Given end node $t$, every solution $\sol$ must satisfy $end(\sol) = t$.
    \item[= \emph{False}] For any solution $\sol$, $end(\sol)$ can be any node in $V$.
\end{itemize}

\medskip
\noindent \textbf{\emph{circuit}: bool}\\
The \emph{circuit} attribute describes whether solutions are closed walks and the traveler returns to its start node at the end of the walk or not.
\begin{itemize}[align=left,labelwidth=\widthof{= \emph{False}},leftmargin=\labelwidth + \labelsep]
    \item[= \emph{True}] Every solution $\sol$ satisfies $start(\sol) = end(\sol)$.
    \item[= \emph{False}] There may exist a solution $\sol$ such that $start(\sol) \ne end(\sol)$.
\end{itemize}

\medskip
\noindent \textbf{\emph{graphtype}: string}\\
The \emph{graphtype} attribute defines characteristics of the underlying graph. For example, a complete graph is considered in ~\cite{dantzig1954solution}.
\begin{itemize}[align=left,labelwidth=\widthof{= \emph{strongly connected}},leftmargin=\labelwidth + \labelsep]
    \item[= \emph{complete}] For all $u,v\in V$: $\{u,v\}\in E$ if $G$ is undirected, or $(u,v)\in E$  and  $(v,u)\in E$ if $G$ is directed.
       \item[= \emph{strongly connected}] For all $u,v\in V$, there exists a directed walk $(u, e_1, ..., e_k, v)$ where $e_i \in E$ for all $i \in \{1, ..., k\}$.
    \item[= \emph{planar}] The graph is planar, \ie it can be embedded in the Euclidean plane without edge crossings, equivalently, it neither contains a $K_5$-minor nor a $K_{3,3}$-minor.
    \item[= \emph{tree(b)}] The graph is a tree with $b$ leaves. Without the parameter $b$, the number of leaves is arbitrary.
    \item[= \emph{path}] The graph is a path, \ie a tree with only 2 leaves, so is equivalent to \emph{tree($2$)}.
    \item[= \emph{cycle}] The graph is a circular graph, \ie consists of a simple cycle.
    \item[= \emph{binary tree}] The graph is a binary tree.
\end{itemize}
If none of the above attribute values are specified, the graph can be arbitrary.

\medskip
\noindent \textbf{\emph{edgetype}: string}\\
The \emph{edgetype} attribute defines properties of edges present in the graph. Commonly occurring values are undirected, directed, or bidirected. In a bidirected graph as, for example, considered in~\cite{papadimitriou2006approximability}, edges are directed, but whenever an edge from some node $u$ to some node $v$ exists, the return edge from $v$ to $u$ must also exist. This enables the definition of problems where it is always guaranteed that if one can travel directly between two nodes, one can travel in both directions, but the edge costs depend on the traversal direction, such as for example when cycling up or down a hill.
\begin{itemize}[align=left,labelwidth=\widthof{= \emph{undirected}},leftmargin=\labelwidth + \labelsep]
    \item[= \emph{undirected}] Each edge $\{u,v\}$ is an unordered two-element subset of $V$.
    \item[= \emph{directed}] Each edge $(u,v)$ is an ordered pair in $V\times V$.
    \item[= \emph{bidirected}] Each edge $(u,v)$ is an ordered pair in $V\times V$ and $(v,u)\in E$.
\end{itemize}

\medskip
\noindent \textbf{\emph{precedences}: string}\\
The \emph{precedences} attribute describes whether precedence constraints exist between nodes in the graph, which specify whether certain cities have to be visited before others, see, for example,~\cite{balas1995precedence, koehler2021cable}.
\begin{itemize}[align=left,labelwidth=\widthof{= \emph{arbitrary}},leftmargin=\labelwidth + \labelsep]
    \item[= \emph{atomic}] There exists a set of atomic precedence constraints $A\subset V\times V$. Any solution $\sol$ must satisfy that for all $(u,v)\in A$, if there exists $j$ such that $\sol_V(j)=v$, then there must exist $i$ such that $i<j$ and $\sol_V(i)=u$.
    \item[= \emph{arbitrary}] There exist constraints defined in some constraint language that are specified outside the definition scheme. These constraints should be described in the $\oplus$ extension of the \tico definition. 
\end{itemize}
The values are currently stated for visited nodes rather than traversed nodes in a solution. However, they can be  extended to traversed nodes.

\medskip
\noindent \textbf{\emph{cluster}: string}\\
The \emph{cluster} attribute restricts the visiting order of nodes in the graph. Nodes are grouped into clusters $\{P_1,\dots,P_k\}$ and all nodes within each cluster must be visited consecutively in a solution sequence. The cluster can be partitions, where each node is in exactly one cluster~\cite{laporte1997tabu}, or covers~\cite{hansknecht2021dynamic}, where nodes may be in more than one cluster. Two optional parameters in parentheses can follow the attribute value to define further restrictions on the order in which the nodes are to be visited. The first parameter \texttt{order} restricts a possible visiting order between the clusters and the second parameter \texttt{sequence} restricts the visiting sequence within a cluster, often specifying entry and exit nodes in a cluster. 
 
 \medskip
\begin{itemize}[align=left,labelwidth=\widthof{= \emph{partition}(\texttt{order}, \texttt{sequence})},leftmargin=\labelwidth + \labelsep]
    \item[= \emph{partition}(\texttt{order}, \texttt{sequence})]
    For all $v\in V$, there exists a unique $a\in \{1,\dots,k\}$ such that $v\in P_a$ and any solution $\sol$ must satisfy that if $i<j<l$ and $\sol_V(i)\in P_a$ and $\sol_V(l)\in P_a$ then $\sol_V(j)\in P_a$.
    \item[= \emph{cover(\texttt{order}, \texttt{sequence})}] For all $v\in V$, there exists at least one $a\in \{1,\dots,k\}$ such that $v\in P_a$ and any solution $\sol$ must satisfy that if $i<j<l$ and $\sol_V(i)\in P_a$ and $\sol_V(l)\in P_a$ then $\sol_V(j)\in P_a$.
\end{itemize}

\medskip
\noindent
For the optional \texttt{order} parameter, the value \emph{ordered} is defined:\footnote{The value \emph{unordered} is the default value and the parameter can be omitted in this case.}

\medskip
\begin{itemize}[align=left,labelwidth=\widthof{\emph{ordered}},leftmargin=\labelwidth + \labelsep]%
    \item[\emph{ordered}] For any solution $\sol$ and for all $i<j$ with $\sol_V(i)\in P_a$ and $\sol_V(j)\in P_b$, it must hold that $a \leq b$.
\end{itemize} 

\medskip
\noindent    
For the optional \texttt{sequence} parameter, the following values are defined:

\medskip
\begin{itemize}[align=left,labelwidth=\widthof{\emph{terminals}},leftmargin=\labelwidth + \labelsep]%
\item[\emph{start}] Given start nodes $\{s_1,\dots,s_k\} \subset V$, any solution $\sol$ must satisfy that $ s_j = \sol_V (\min\{i : \sol_V(i)\in P_j\}) $ for all $j\in\{1,\dots,k\}$.
\item[\emph{startend}] Given start nodes $\{s_1,\dots,s_k\} \subset V$ and end nodes $\{t_1,\dots,t_k\} \subset V$, any solution $\sol$ must satisfy that 
$s_j = \sol_V (\min\{i : \sol_V(i)\in P_j\})$ and $t_j = \sol_V (\max\{i : \sol_V(i)\in P_j\})$
for all $j\in \{1,\dots,k\}$.
\item[\emph{terminals}] Given a set of potential start and end nodes $T_j\subseteq P_j$ for each $j\in \{1,\dots,k\}$, any solution $\sol$ must satisfy that $\{\sol_V (\min\{i : \sol_V(i)\in P_j\}), \sol_V (\max\{i : \sol_V(i)\in P_j\})\}\subseteq T_j$.
\end{itemize}

\medskip
\noindent
As before, the values can be extended to traversed nodes.

\subsection{The Costs Field \texorpdfstring{$\delta$}{δ}}
\label{section:costfield}

Several cost functions can be defined for a problem in the $\delta$-field. They are identified by a unique name and specified by domain and range. The syntax is uniformly defined for the longhand and the shorthand notation, recall Figures~\ref{fig:ebnf-long} and~\ref{fig:ebnf-short}:
\[
\textit{cost function}   \Leftarrow  \textit{name ``:'' domain ``$\mapsto$'' range \{ ``,'' attribute \} }
\]
\noindent
The \emph{domain}  of the cost function is usually taken from the set of edges $E$ (edge costs) or nodes $V$ (node costs) in the graph. An example of using node costs can be found in~\cite{laporte1990selective}. The quadratic TSP as described in~\cite{fischer2013symmetric} uses the pair of edges $E\times E$. The \emph{range} of the cost function can be any subset of the real numbers $\mathbb{R}$. The subsequent attributes are used to further characterize a cost function in the $\delta$-field.

\medskip
\noindent \textbf{\emph{property}: string}\\
The \emph{property} attribute is defined for cost functions with domain $E$. For example, edge costs may satisfy the triangle inequality~\cite{ding2007two}, be determined by a pseudometric~\cite{glebov20175}, a metric~\cite{karlin2021slightly}, a graphic metric~\cite{momke2011approximating}, a planar metric~\cite{Kle2005}, an induced metric subspace of a planar metric~\cite{Kle2006}, or a euclidean metric~\cite{arora1996polynomial}. Due to the non-uniform definitions of pseudometric and quasimetric that we observed in the literature, we introduce parameters to further characterize cost functions. Any premetric properties are explicitly listed in "()". If multiple properties apply, the most specific one should be listed. For example,  \emph{euclid} is more specific than  \emph{metric}, \emph{planar} is more specific than \emph{graphic}, and  \emph{graphic} implies \emph{triangle}.

\begin{itemize}[align=left,labelwidth=\widthof{= \emph{euclidean fixed dim}},leftmargin=\labelwidth + \labelsep]
\item[= (\texttt{params})] The edge cost function satisfies some attributes of a metric that are specified by parameters in "()".\footnote{In particular, self loops have costs 0, \ie if $(u,u) \in E$ then  $c(u,u) = 0$  for all $u \in V$, and edge costs are non-negative. TSP with self loops with costs strictly larger than 0 or TSP with negative edge costs, do not define a premetric.} 
\item[= \emph{metric}] The edge cost function satisfies the parameters \emph{symmetry}, \emph{identity}, and \emph{triangle}.
\item[= \emph{graphic}] The edge cost function corresponds to the number of edges of the shortest path between two nodes in an unweighted graph. Consequently, it is a metric. 
\item[= \emph{planar}] The edge cost function corresponds to the cost of the shortest path between two nodes in a planar graph with non-negative weights. Consequently, it is a metric.
\item[= \emph{subset planar}] The edge cost function is a metric. The points of this metric correspond to a subset of the nodes of a planar graph. The distances between the points are the costs of the shortest paths using all nodes and edges, which must have non-negative lengths, in the entire planar graph.

\item[= \emph{euclidean\label{def:euclidean}}] The edge cost function is determined by a euclidean metric.
\item[= \emph{euclidean fixed dim}] The edge cost function is determined by a fixed-dimensional euclidean metric.
\item[= \emph{euclidean plane}] The edge cost function is determined by a euclidean metric in the plane.
\item[= \emph{grid(m,n)}] The edge cost function is determined by a euclidean metric induced by $m \times n$ grid points of an equidistant two-dimensional array.
\end{itemize}

\medskip
\noindent
As possible parameters \texttt{params}, a subset of the following can occur in "()":
\begin{itemize}[align=left,labelwidth=\widthof{\emph{symmetric}},leftmargin=\labelwidth + \labelsep]%
\item[\emph{identity}] For all  $(u,v)\in E$, it holds that $c(u,v)>0$ if and only if $u\neq v$.
\item[\emph{triangle}] For all $u,v,w \in V$, it holds that $c(u,v)\leq c(u,w) + c(w,v)$.
\item[\emph{$\alpha$-triangle}] For all $u,v,w \in V$, it holds that  
$c(u,v) \leq \alpha \cdot ( c(u,w) + c(w,v))$.
\item[\emph{symmetric}] For all $(u,v) \in E$, it holds that $(v,u)\in E$ and $c(u,v) = c(v,u)$.\footnote{Note that the value \emph{undirected} of the attribute \emph{edgetype} implies the parameter \emph{symmetric} of the cost function. If symmetry is not required, then asymmetric costs are permitted.}

\end{itemize}

\medskip
\noindent
For cost functions with domain $E$ or domain $V$ or a combinations of both, we define the following two attributes:

\medskip
\noindent
\textbf{\emph{partial}: bool}\\
The \emph{partial} attribute describes whether only some of the edge or node cost are incurred on a tour. For example, if the cost corresponds to the availability of a product in certain cities, then the traveler may only buy some of the available product, see for example~\cite{manerba2017traveling,bontoux2008ant}. 

\begin{itemize}[align=left,labelwidth=\widthof{= \emph{False}},leftmargin=\labelwidth + \labelsep]
    \item[= \emph{True}] A tour only incurs partial edge or node costs.
    \item[= \emph{False}] A tour always incurs the full cost values.
\end{itemize}

\medskip
\noindent \textbf{\emph{temporal}: string}\\
The \emph{temporal} attribute describes whether the cost function has a temporal dependency and may thus change over time. Typical examples are cost functions that depend on the cost of nodes or edges~\cite{hansknecht2021dynamic,malandraki1996restricted} or on the number of edges~\cite{gouveia1995classification,picard1978time} of the prefix of the walk at the current node. Such cost functions may also only change a fixed number of times, \eg~\cite{broden2004online}. In the Kinetic-based TSP or Moving-target TSP, \eg~\cite{helvig1998moving}, nodes move with constant speed in a fixed direction and edge costs between nodes change continuously.

\begin{itemize}[align=left,labelwidth=\widthof{= \emph{costzone}$(k)$},leftmargin=\labelwidth + \labelsep]%
    \item[= \emph{time}]  For all $u, v \in V$, $c(u,v)$ or $c(u)$ is a function of the arrival time of the prefix of the walk at node $u$.
    \item[= \emph{position}] For all $u,v \in V$, $c(u,v)$ or $c(u)$ depends on the number of edges and/or nodes of the prefix of the walk at node $u$.  
    \item[= \emph{costzone}$(k)$] The cost function satisfies the conditions of value \emph{time} and changes at most $k-1$ times.
    \item[= \emph{poszone}$(k)$] The cost function satisfies the conditions of value \emph{position} and changes at most $k-1$ times.
    \item[= \emph{waiting}] When travelers arrive at a node $u$, they may wait before they visit the node $u$.
    \item[= \emph{kinetic}]  Each node moves with constant speed in a fixed direction and edge costs between nodes change continuously.
\end{itemize}

\subsection{The Objectives Field \texorpdfstring{$\epsilon$}{ε}}

The objectives field defines the objective(s) in the problem, \ie which costs are to be bounded, maximized or minimized and be written as an arbitrary string. Suppose $c$ is a cost function specified in the $\delta$-field, then $c(\sol)$ denotes the sum of costs of the walk $\sol$ and $\compl{c}(\sol)$ denotes the complement of these costs. 

We will use simple arithmetic expressions in these strings, which are mostly self-explanatory. For example, if two cost models $c_1$ and $c_2$ were defined in the $\delta$-field and the objective is to minimize their sum,  we write $\min c_1(\sol) + c_2(\sol)$ in the $\epsilon$-field. If the objective is to maximize their sum, we write $\max c_1(\sol) + c_2(\sol)$ in the $\epsilon$-field. If there is a constraint on the maximum value of the function over cost models, \ie an upper bound $b$, we write  $c_1(\sol) + c_2(\sol) \leq b$. In case of a lower bound  $b$, we write for example $c_1(\sol) + c_2(\sol) \geq b$.

\section{Review of Approximability Results}
\label{section.approximability}

In this section, we review results for TSP variants known as the the "standard" TSP,  Path TSP, Bottleneck TSP, Maximum Scatter TSP,  Generalized TSP, Clustered TSP, Traveling Purchaser Problem, Profitable Tour Problem, Quota TSP, Prize-Collecting TSP, Orienteering Problem, Time-dependent TSP, TSP with Time Windows, and the Orienteering Problem with Time Windows. After a brief informal description of the problem variant, we present an overview of approximability and inapproximability results where we limit ourselves to only peer reviewed results and focus on the best currently known bounds. We selected a paper for inclusion in the review if the paper either uses the name of the problem variant or provides a definition that is very close to other publications using the name. For hybrid TSP variants, we only included the Orienteering Problem with Time Windows due to the extensive body of results. Further approximability results exist for other hybrid variants. For each reference from the literature, we provide a \tico definition using the shorthand notation. The definition makes it possible to very compactly and clearly summarize the results in table form and it clarifies the restrictions under which a result holds. Note that we do not review  algorithms running only in quasi-polynomial time, but only polynomial time results. However, occasionally in the text, we point to further results outside the scope of this paper.

In the presentation of the approximability results, we use an asymptotic notation for upper and lower bounds. This typically includes big-O and little-o for upper bounds, whereas big-$\Omega$ and little-$\omega$ is used in lower bounds. Informally speaking, the difference between big-O and little-o is similar to the difference between $\leq$ and $<$, whereas the difference between big-$\Omega$ and little-$\omega$ is similar to the difference between $\geq$ and $>$. 

More formally, for a function $f: \mathbb{Z}_{\ge 0} \to \mathbb{R}$ we define:
\begin{itemize}
\item $O(f(n)) := \{ g: \mathbb{Z}_{\ge 0} \to \mathbb{R}\, : \exists c, n_0 > 0, \forall n \ge n_0: |g(n)| \leq c \cdot |f(n)| \}$
\begin{center}
 is the set of functions that grow \textbf{at most as fast} as $f$.
\end{center}
\item $\Omega(f(n)) := \{ g: \mathbb{Z}_{\ge 0} \to \mathbb{R}\, : \exists c, n_0 > 0, \forall n \ge n_0: |f(n)| \leq c \cdot |g(n)| \}$
\begin{center}
 is the set of functions that grow \textbf{at least as fast} as $f$.
\end{center}

\item $o(f(n)) := \{ g: \mathbb{Z}_{\ge 0} \to \mathbb{R}\, : \forall c > 0, \exists n_0 > 0, \forall n \ge n_0: |g(n)| \leq c \cdot |f(n)| \}$
\begin{center}
is the set of functions that grow \textbf{strictly slower} than $f$.
\end{center}
\item $\omega(f(n)) := \{ g: \mathbb{Z}_{\ge 0} \to \mathbb{R}\, : \forall c > 0, \exists n_0 > 0, \forall n \ge n_0: |f(n)| \leq c \cdot |g(n)| \}$
\begin{center}
is the set of functions that grow \textbf{strictly faster} than $f$.
\end{center}
\end{itemize}
In general, we use $\epsilon$ to denote any arbitrarily small positive real number. Though the asymptotic notation defines sets, we also use it in mathematical expressions and equations. The former case describes a transformed set by means of composition, whereas a statement is considered true if it holds for at least one function in the corresponding set. Furthermore, when a result holds for non-negative edge costs, it can be generalized to also include negative edge costs unless there is a directed cycle with negative cost. This is due to \emph{Johnson's trick}~\cite{Johnson77}: In the absence of negative directed cycles, there exist node potentials $p: V \to \mathbb{R}$ such that, for each edge $e=(v,w)$ with cost $c(e)$, the reduced cost $c_p(e) := c(e) + p(v) - p(w)$ is non-negative. When we add the reduced costs along a directed cycle, the node potentials cancel out, so that the cost of any cycle is the same as its reduced cost. Such node potentials can be found in polynomial time using the Bellman-Ford algorithm.

From a complexity point of view, the encoding of the numbers in the input matters. However, reals can be approximated by rationals to any desired degree of accuracy such that the influence on approximation factors becomes insignificant. We therefore use $\mathbb{R}$ for the domain of cost functions unless further restrictions apply.

\subsection{The ``standard'' TSP}
\label{sec.standardtsp}
We consider the ``standard'' TSP to be the problem of finding a shortest walk through a graph traversing all cities and returning to the origin city at the end. The edge costs may be assumed to be asymmetric~\cite{syslo1980generalizations} or some restrictions may be placed on the edge costs, for example that they are euclidean~\cite{arora1996polynomial}. Additionally, the underlying graph may be considered to be complete \cite{syslo1980generalizations} or incomplete \cite{bellmore1974transformation}.

It is a well-known fact that no polynomial time $\rho$-approximation algorithm exists for the ``standard'' TSP in which each city is to be traversed exactly once, where $\rho$ is any polynomial time computable function, unless P=NP. We indicate this with $\infty$ as lower bound (see the first row of Table~\ref{table.standardTSP}). As most extensions of the TSP can be reduced to a ``standard'' TSP, this inapproximability result generalizes to most of the extensions presented in this section. Furthermore, this result means that approximability results can only exist for variants of the ``standard'' TSP, which simplify the problem by applying further restrictions on the cost function, or by relaxing restrictions on the tour.

An overview of approximability and inapproximability results is given in Table~\ref{table.standardTSP}.

\begin{table}[htp] \begin{tabular}{p{\textwidth}} $ $ \end{tabular}\vspace{-7ex}
\[
\begin{array}{lclcr}
\textbf{Variant} & \textbf{Lower Bound} & & \textbf{Upper Bound} & \\ \hline \hline
\langle =1 \,|\, =1 \,|\, \text{circuit; undirected} \,| & \infty & \text{\cite{Karp1972}} & & \\ 
\hspace{5mm} c : E\mapsto \mathbb{R} \,|\, \min c(\sol)\rangle \\\hline

\langle =1 \,|\, \geq 1 \,|\, \text{circuit; strongly connected; directed} \,| & \frac{75}{74} & \text{\cite{KLS2015}} \confirmed & 22 + \varepsilon &\text{\cite{TV2020}} \confirmed\\
\hspace{5mm} c : E\mapsto\mathbb{R}_{\geq 0}\,|\, \min c(\sol)\rangle\\ \hline

\langle =1 \,|\, =1 \,|\, \text{circuit; complete; undirected} \,|& \frac{123}{122} & \text{\cite{KLS2015}} \confirmed & \frac{3}{2}-10^{-36} & \text{\cite{karlin2021slightly}} \unconfirmed\\
\hspace{5mm} c : E\mapsto\mathbb{R}_{\geq 0}, \text{metric}\,|\, \min c(\sol) \rangle&&&&\\ \hline

\langle =1 \,|\, =1 \,|\, \text{circuit; complete; undirected} \,| & & & \frac{7}{5} & \text{\cite{seboe2014}} \confirmed\\
\hspace{5mm}  c : E\mapsto\mathbb{R}_{\geq 0}, \text{graphic} \,|\, \min c(\sol) \rangle&&&&\\\hline

\langle =1 \,|\, =1 \,|\, \text{circuit; complete; undirected} \,| & &  & 1+ \varepsilon & \text{\cite{arora1996polynomial,Mit1999}} \unconfirmed\\
\hspace{5mm}  c : E\mapsto\mathbb{R}_{\geq 0}, \text{euclid} \,|\, \min c(\sol) \rangle&&&&\\\hline

\langle =1 \,|\, =1 \,|\, \text{circuit; complete; undirected} \,|& &  & 1+ \varepsilon &\text{\cite{Kle2005}} \confirmed\\
\hspace{5mm} c : E\mapsto\mathbb{R}_{\geq 0}, \text{planar} \,|\, \min c(\sol) \rangle&&&& \\ \hline

\langle =1 \,|\, =1 \,|\, \text{circuit; complete; undirected} \,|& &  & 1+ \varepsilon &\text{\cite{Kle2006}} \confirmed\\
\hspace{5mm} c : E\mapsto\mathbb{R}_{\geq 0}, \text{subset planar} \,|\, \min c(\sol) \rangle&&&& \\ \hline \hline

\end{array}
\]
    \caption{Approximability and inapproximability results for the ``standard'' TSP. Note that $=1$ in the $\beta$ field can be replaced by $\geq 1$ in all variants with a cost function satisfying the triangle inequality, see Fact~\ref{fact}.}
    \label{table.standardTSP}
\end{table}

Most importantly, the ``standard'' TSP with metric edge costs and on a complete graph can at best be approximated with a ratio of $123/122$~\cite{KLS2015}. The first algorithm giving an upper bound better than $3/2$ for the same variant was presented~\cite{karlin2021slightly}, which is considered a major breakthrough. If more restrictions are imposed on the values of the edge costs, so required to be euclidean or given by a graphic metric of a planar graph, then the ``standard'' TSP can be approximated arbitrarily closely \ie with a ratio of $1+\epsilon$ in polynomial time. Finally, note that the following holds:

\begin{fact} \label{fact}

The following two \tico variant definitions are equivalent:
\begin{align}
& \langle =1\,|\,=1\,|\,\text{circuit; complete; undirected}\,|\, c :  E \mapsto \mathbb{R}_{\geq 0}, \text{ (triangle)} \,|\,  \min c(\sol) \rangle
\\
& \langle =1\,|\,\geq 1\,|\,\text{circuit; undirected}\,|\, c :  E \mapsto \mathbb{R}_{\geq 0} \,|\, \min c(\sol) \rangle
\end{align}
Thus, any results that apply to one, also transfer to the other. 
\end{fact}

However, there is a subtle difference between \emph{planar} and \emph{subset planar} metric spaces in the last two entries of Table~\ref{table.standardTSP}. The first of the two is equivalent to
$\langle =1\,|\,\geq 1\,|\,\text{circuit; planar; undirected}\,|\, c :  E \mapsto \mathbb{R}_{\geq 0} \,|\, \min c(\sol) \rangle$,
whereas the latter is equivalent to 
$\langle =1\,|\,\geq d(v) \in \{0,1\}\,|\,\text{circuit; planar; undirected}\,|\, c :  E \mapsto \mathbb{R}_{\geq 0} \,|\, \min c(\sol) \rangle$
for some characteristic function $d$ indicating some subset $S$ of the nodes. The latter problem is also called planar Subset TSP or planar Steiner TSP. Note that the graph induced by $S$ generally does not yield the same shortest path distances and it might not be possible to generate the metric by a planar graph on $S$.

\subsection{Path TSP}
\label{sec.path}

In the Path TSP, also denoted as the $s-t$ Path TSP, the goal is to find the shortest Hamiltonian path between two nodes $s$ and $t$ in the graph rather than a cycle traversing all nodes. For a long time, there was a gap between the best known approximation bounds for the circuit version of the TSP compared to the path version of the TSP, see Table~\ref{table.pathTSPvariants} for three variants of the Path TSP that have been studied. However, in 2020, it was shown~\cite{traub2020reducing} that any $\alpha$-approximation for a standard TSP can be used to obtain an $(\alpha+\epsilon)$-approximation for the corresponding path variant. To this end, the $\alpha$-approximation is called on instances defined on subgraphs from the instance for Path TSP. Hence, for any standard TSP with non-negative symmetric edge costs having some some property $\mathbf{p}$ that is preserved under taking subgraphs
\[
\langle =1 \,|\, \geq 1 \,|\, \text{circuit; complete; undirected} \,|\, c :  E \mapsto \mathbb{R}_{\geq 0}, \mathbf{p} \,|\, \min c(\sol)\rangle 
\]
that admits an $\alpha$-approximation, we have an $(\alpha+\epsilon)$-approximation for the Path TSP where the non-negative symmetric edge costs also have property $\mathbf{p}$
\[
\langle =1 \,|\, \geq 1 \,|\, \mbox{start; end; complete; undirected} \,|\, c :  E \mapsto \mathbb{R}_{\geq 0}, \mathbf{p} \,|\, \min c(\sol) \rangle
\] 
for any arbitrarily small constant $\varepsilon > 0$. The currently best known upper approximation bounds for the Path TSP can thus be obtained from the upper approximation bounds for the standard TSP by setting the \textit{circuit} attribute to False, the \textit{start} and \textit{end} attributes to True and by adding $+ \epsilon$ to the approximation bound.

\begin{table}[htp] \begin{tabular}{p{\textwidth}} $ $ \end{tabular}\vspace{-7ex}
\[
\begin{array}{lr}
\textbf{Variant} & \textbf{Reference} \\\hline \hline

\langle =1 \, | \, =1\, | \,  \mbox{start; end; complete; undirected} \, | \, c : E \mapsto \mathbb{R}_{\geq 0}, \mbox{metric}\, | \,\min c(\sol)\rangle& \text{\cite{an2015improving,traub2020reducing,seboe2013,seboeVanZuylen2016}} \confirmed \text{\cite{zenklusen2019}}\unconfirmed\\ \hline

\langle =1 \, | \, =1\, | \,  \mbox{start; end; complete; directed} \, |  \, c : E\mapsto \mathbb{R}_{\geq 0}, \mbox{triangle} \, | \, \min c(\sol)\rangle&  \text{\cite{koehne2020asymmetric}}\confirmed \text{\cite{feige2007improved,svensson2020constant}}\unconfirmed  \\ \hline

\langle =1 \, | \, \geq 1\, | \,  \mbox{start; end; undirected} \, | \, c : E\mapsto \{1\}\, | \, \min c(\sol)\rangle & \text{\cite{seboe2014,traub2020beating,traub2020reducing}}\confirmed   \text{\cite{mucha2014approximation,an2015improving,moemke2016removing}}\unconfirmed \\ \hline \hline
\end{array}
\]
    \caption{Variants of the Path TSP.}
    \label{table.pathTSPvariants}
\end{table}

Various hybrid variants of the Path TSP are also studied in the literature. An example is the Metric Many-visit Path TSP~\cite{rothkopf1966traveling}, in which each city needs to be traversed a pre-specified number of times.  It can be defined in \tico as
\[\langle =1 \,|\, d_v \,|\, \mbox{start; end; complete; undirected} \,|\, c :  E \mapsto \mathbb{R}_{\geq 0}, \mbox{metric} \,|\, \min c(\sol) \rangle
\]
A polynomial time approximation algorithm with an approximation ratio of 1.5 for this problem was devised in~\cite{berczi2020}.

The Path TSP is also denoted in the literature as the Messenger Problem~\cite{gutin2006traveling,menger1932botenproblem}, the Wandering Salesman Problem~\cite{djang1993wandering,papadimitriou1984two}, or the Minimum Hamiltonian Path problem (MHP)~\cite{hammar2002approximation}.

\subsection{Bottleneck TSP}
\label{sec.bottleneck}
The Bottleneck TSP differs from the ``standard'' TSP only in the objective function \cite{garfinkel1978bottleneck}. Unlike in the ``standard'' TSP, where the objective is to minimize the sum of edge costs along the solution path, the objective of the bottleneck TSP is to minimize the maximum edge cost along the path. Furthermore, the tour has to traverse all nodes, but is not required to be a circuit. An overview of approximability and inapproximability results is given in Table~\ref{table.bottleneck}.
\begin{table}[htp] \begin{tabular}{p{\textwidth}} $ $ \end{tabular}\vspace{-7ex}
\[
\begin{array}{lcrcr}
\textbf{Variant} & \textbf{Lower Bound} & & \textbf{Upper Bound} & \\
\hline \hline
\langle =1 \,|\, = 1 \,|\, \text{circuit; complete; undirected} \,| & \omega(1) & \text{\cite{PR1984}} & &  \\
\hspace{5mm} c : E\mapsto\mathbb{R} \,|\, \min \max \{c(e) : e\in E_{\sol}\}\rangle\\
\hline
\langle =1 \,|\, = 1 \,|\, \text{circuit; complete; undirected} \,| & 2 & \text{\cite{PR1984}} & 2 & \text{\cite{PR1984}} \\
\hspace{5mm} c : E\mapsto\mathbb{R}_{\geq 0}, \text{(triangle)} \,|&&&&\\
\hspace{5mm} \min \max \{c(e) : e\in E_{\sol}\}\rangle\\\hline
\langle =1 \,|\, =1 \,|\, \text{circuit; complete; directed} \,| & & & \lceil n/2 \rceil & \text{\cite{larusic2014asymmetric}} \\ 
\hspace{5mm} c : E\mapsto \mathbb{R}_{\geq 0}, \text{(triangle)} \,| &&&&\\ 
\hspace{5mm} \min \max \{c(e) : e\in E_{\sol}\}\rangle \\
\hline
\langle =1 \,|\, =1 \,|\, \text{circuit; complete; bidirected} \,| & & & \tfrac{\tau}{\tau-1}(2\tau^{\lceil n/2\rceil -1}-\tau^{\lceil n/2\rceil-2}-1) & \text{\cite{larusic2014asymmetric}} \\ 
\hspace{5mm} c : E\mapsto \mathbb{R}_{\geq 0}, \text{($\tau$-triangle)} \,| &&&&\\ 
\hspace{5mm} \min \max \{c(e) : e\in E_{\sol}\}\rangle \\
\hline
\langle =1 \,|\, =1 \,|\, \text{circuit; complete; bidirected} \,| & & & \tfrac{\lambda}{\lambda - 1}(\lambda^{\lceil n/2\rceil - 1} + \lambda -2) & \text{\cite{larusic2014asymmetric}} \\ 
\hspace{5mm} c : E\mapsto \mathbb{R}_{\geq 0}, \text{($\lambda$-triangle)} \,| &&&&\\ 
\hspace{5mm} \min \max \{c(e) : e\in E_{\sol}\}\rangle \\
\hline
\langle =1 \,|\, =1 \,|\, \text{circuit; directed} \,| & 2 & \text{\cite{AKS2021}} & O(\log n / \log\log n) & \text{\cite{AKS2021}} \\ 
\hspace{5mm} c : E\mapsto \mathbb{R}_{\geq 0}, \text{(triangle)} \,| &&&&\\ 
\hspace{5mm} \min \max \{c(e) : e\in E_{\sol}\}\rangle \\
\hline
\langle =1 \,|\, =1 \,|\, \text{complete; bidirected} \,| & & & 2 + \gamma & \text{\cite{kao2009approximation}} \\ 
\hspace{5mm} c : E\mapsto \mathbb{R}_{\geq 0} \,| &&&&\\ 
\hspace{5mm} \min \max \{c(e) : e\in E_{\sol}\}\rangle^{\oplus 1} \\
\hline \hline
\end{array}
\]
\begin{description}[leftmargin=0.6cm,labelindent=0cm]
\item[$\oplus 1$:] 
    \begin{itemize}
        \item[--] Each node $n_i$ is specified by a pair of numbers $(\alpha_i, \beta_i)$.
        \item[--] The cost function is then given by $c(n_i, n_j) = d(\alpha_i, \beta_j)$ with $d(a,b) = 
              \int_{a}^{b}f(x) dx$ if $b\geq a$ and $d(a,b) = \int_{\beta_j}^{\alpha_i}g(x) dx$ if $b < a$,
        where the functions $f(x)$ and $g(x)$ are integrable and satisfy $f(x), g(x) \geq 0$ for all $x$. 
        \item[--] The constant $\gamma$ satisfies $\gamma \geq \tfrac{f(x)}{g(x)} \geq \tfrac{1}{\gamma}$ for all $x$.
    \end{itemize}
\end{description}
    \caption{Approximability and inapproximability results for the Bottleneck TSP. The constant $\tau$ satisfies $\tau > 1$, the constant $\lambda$ satisfies $\lambda < 1$, and $\gamma$ is a measure of the asymmetry of the problem, see $\oplus 1$ for details.}
    \label{table.bottleneck}
\end{table}
The Bottleneck TSP is sometimes also called the MinMax TSP \cite{kozma2015ptas}. A closely related variant is the Maximum Scatter TSP, where the objective is to maximize the minimum cost of an edge on the tour, see Section~\ref{sec.maximumscatter}.

\subsection{Maximum Scatter TSP}
\label{sec.maximumscatter}
The Maximum Scatter TSP \cite{arkin99}, similarly to the previously discussed Bottleneck TSP, differs from the ``standard'' TSP in the objective function. The objective here is to maximize the minimum edge cost along the path. Furthermore, the tour has to traverse all nodes, but is not required to be a circuit. An overview of approximability and inapproximability results is given in Table~\ref{table.maximumscatter}.

\begin{table}[htp] \begin{tabular}{p{\textwidth}} $ $ \end{tabular}\vspace{-7ex}
\[
\begin{array}{lcrcr}
\textbf{Variant} & \textbf{Lower Bound} & & \textbf{Upper Bound} & \\
\hline \hline
\langle =1 \,|\, = 1 \,|\, \text{circuit; complete; undirected} \,| & \omega(1) & \text{\cite{arkin99}} \confirmed& &  \\
\hspace{5mm} c : E\mapsto \mathbb{R}_{> 0} \,|\, \max \min \{c(e) : e\in E_{\sol}\}\rangle\\ \hline
\langle =1 \,|\, = 1 \,|\, \text{circuit; complete; undirected} \,| & 2 & \text{\cite{arkin99}} \confirmed & 2 & \text{\cite{arkin99}}  \confirmed\\
\hspace{5mm} c : E\mapsto \mathbb{R}_{\geq 0}, \text{(triangle)} \,|\, \max \min \{c(e) : e\in E_{\sol}\}\rangle\\ \hline
\langle =1 \,|\, = 1 \,|\, \text{complete; undirected} \,| & 2 & \text{\cite{arkin99}}\confirmed & 2 & \text{\cite{arkin99}} \confirmed \\
\hspace{5mm} c : E\mapsto \mathbb{R}_{\geq 0}, \text{(triangle)} \,|\, \max \min \{c(e) : e\in E_{\sol}\}\rangle\\ \hline
\langle =1 \,|\, = 1 \,|\, \text{circuit; complete; undirected;} \,| & & & 1+ o(1) & \text{\cite{hoffmann2017maximum}} \confirmed \\
\hspace{5mm} c : E\mapsto \mathbb{R}_{\geq 0}, \text{grid(m,n)} \,|\, \max \min \{c(e) : e\in E_{\sol}\}\rangle & & & &\\
\hline \hline
\end{array}
\]

    \caption{Approximability and inapproximability results for the Maximum Scatter TSP. When $m=n$ or $m=2$, the upper bound in~\cite{hoffmann2017maximum} is 1.}
    \label{table.maximumscatter}
\end{table}

The Maximum Scatter TSP is, in general, NP-hard and no constant-factor approximation algorithm for it exists, unless P=NP. Due to the restrictions on the cost functions, the approximability results for the Bottleneck TSP can not directly be translated to approximability results for the Maximum Scatter TSP. As for the Bottleneck TSP, a best-possible polynomial time approximation algorithm exists for the Maximum Scatter TSP achieving an approximability ratio of 2. When the nodes lie on a line or a circle in the euclidean plane, and the cost function is given by the euclidean distance between the nodes, an exact linear time algorithm exists~\cite{arkin99}. For the extension of the Maximum Scatter TSP from direct neighbors to $m$-neighbors, further approximability results exist~\cite{arkin99,chiang2005new}. The Maximum Scatter TSP is sometimes also called the MaxMin TSP \cite{kozma2015ptas}.

\subsection{Generalized TSP}
\label{sec.approx.generalized}

In the Generalized TSP, the nodes of $G$ are grouped into subsets and either exactly one of the nodes or at least one node from each subset has to be traversed. The subsets of nodes may be disjoint and form a partition of $V$, see \eg~\cite{snyder2006random}, or they may overlap and form a cover of $V$, see \eg~\cite{noon1993efficient,noon1988generalized}.
In the variants studied in \cite{fischetti1997branch, noon1993efficient,noon1988generalized,snyder2006random, cosma2024generalized}, the shortest circuit must be found that traverses each subset once. In~\cite{laporte2002some}, each subset must be traversed exactly once, whereas in~\cite{noon1993efficient,noon1988generalized}, each subset must be traversed at least once. Some variants of the Generalized TSP are given in Table~\ref{table.generalized_examples}.

\begin{table}[htp] \begin{tabular}{p{\textwidth}} $ $ \end{tabular}\vspace{-7ex}
\[
\begin{array}{lr}
\textbf{Variant} & \textbf{Reference} \\ \hline\hline

\langle =1 \, | \,  \mbox{$\leq 1$; partition($\geq$ once)}\, | \,  \mbox{circuit; complete; undirected} \, | \, c :  E \mapsto \,\mathbb{R}\, | \, \min c(\sol)\rangle & \text{\cite{fischetti1997branch}} \confirmed\\\hline

\langle =1 \, | \,  \mbox{$\leq 1$, cover($\geq$ once)}\, | \,  \mbox{circuit; complete; directed} \, | \, c :  E\mapsto \mathbb{R}, \mbox{(triangle)}\, | \, \min c(\sol)\rangle & \text{\cite{lien1993transformation}} \confirmed \\\hline

\langle =1 \, | \,  \mbox{$\leq 1$; partition(once)}\, | \,  \mbox{circuit; complete; undirected} \, | \, c :  E\mapsto \mathbb{R}\, | \, \min c(\sol)\rangle & \text{\cite{fischetti1997branch}} \confirmed  \, \text{\cite{laporte2002some}} \unconfirmed\\ \hline

\langle =1 \, | \,  \mbox{$\leq 1$; partition(once)}\, | \,  \mbox{circuit; complete; directed} \, | \, c :  E\mapsto \mathbb{R}\, | \, \min c(\sol)\rangle & \text{\cite{laporte2002some}}  \unconfirmed\\ \hline \hline
\end{array}
\]
    \caption{Variants of the Generalized TSP.}
    \label{table.generalized_examples}
\end{table}
\smallskip  

In \cite{tsiligirides1984heuristic}, the following problem is also referred to as the Generalized TSP, however, it would usually be classified as an Orienteering Problem, cf.\ Section~\ref{sec.orienteering}.
\[
\begin{split}
\langle  =1 \, | \, \leq 1; \mbox{always} \, | \, \mbox{start; end; complete; undirected} \, | \, \\
 \> c: E\mapsto\mathbb{R}, \mbox{(triangle)}; q : V\mapsto\mathbb{R}_{\geq 0} \, |\, c \leq b;\, \max q(\sol) \rangle
\end{split}
\]

When no assumptions are made about the compositions of the subsets, no algorithm giving a constant factor approximation in polynomial time is known. Table~\ref{table.generalized} gives an overview of the polynomial time approximation bounds currently known. Some further polynomial time constant factor approximation results for neighborhoods with specific properties can be found in~\cite{dumitrescu2003approximation} where the subsets correspond to nodes within neighborhoods in the plane and are either connected regions of the same or similar diameter or are disjoint unit disks in the plane. Due to their specificity, we have not included these in the overview here. 
\begin{table}[htp] \begin{tabular}{p{\textwidth}} $ $ \end{tabular}\vspace{-7ex}\vspace{-7ex}
\[
\begin{array}{lcrcr}
\textbf{Variant} & \textbf{Upper Bound} & \\ \hline\hline

\langle =1 \, | \,  \mbox{$\geq 0$; $k$-partition ($\geq$ once)}\, | \mbox{circuit;  undirected} \, |&  O(\log k \log^2n)\, &\text{\cite{garg2000polylogarithmic}} \confirmed \text{\cite{fakcharoenphol2003tight}} \unconfirmed\\
\hspace{5mm}  \, c : E\mapsto \mathbb{R}_{\geq 0}, \mbox{(triangle)}\, | \, \min c(\sol) \rangle & &\\\hline

\langle =1 \, | \,  \mbox{$\leq 1$; $k$-partition (once, $k=O(\log n)$)}\, |& 1+\epsilon \, &\text{\cite{khachay2020complexity}} \confirmed\\
\hspace{5mm}  \mbox{circuit; complete; undirected} \, | \, c : E\mapsto \mathbb{R}_{>0}, \mbox{euclidean}\,| \, \min c(\sol)\rangle^{\oplus 1} & &\\\hline

\langle =1 \, | \,  \mbox{$\leq 1$; $k$-partition (once, $k=n-O(\log n)$)}\, |& 1+\epsilon \, &\text{\cite{khachay2020complexity}} \confirmed\\
\hspace{5mm}  \mbox{circuit; complete; undirected} \, | \, c : E\mapsto\mathbb{R}_{>0}, \mbox{euclidean}\, | \, \min c(\sol)\rangle^{\oplus 1} & &\\ \hline
\langle =1 \, | \,  \mbox{$\leq 1$; $k$-partition (once)}\, |& 1.5 + 8\sqrt{2}+\epsilon \, &\text{\cite{bhattacharya2015approximation}} \unconfirmed\\
\hspace{5mm}  \mbox{circuit; complete; undirected} \, | \, c : E\mapsto\mathbb{R}_{>0}, \mbox{euclidean}\, | \, \min c(\sol)\rangle^{\oplus 1} & &\\ \hline
\langle =1 \, | \,  \mbox{$\geq 0$; $k$-partition ($\geq$ once, $k\geq n-O(\log n)$)}\, |& 22+\epsilon \, &\text{\cite{khachay2022constant}} \unconfirmed\\
\hspace{5mm}  \mbox{circuit; strongly connected; directed} \, | \, c : E\mapsto\mathbb{R}_{\geq 0}\, | \, \min c(\sol)\rangle & &\\ \hline \hline
\end{array}
\]
\begin{description}[leftmargin=0.6cm,labelindent=0cm]
     \item[$\oplus 1$:] The nodes in the problem correspond to a set of points in the euclidean plane and every non-empty cell of the integer grid forms a cluster $P_i$.  
\end{description}
    \caption{Approximability results for the Generalized TSP. Note that the bound given in \cite{garg2000polylogarithmic} was originally determined with $O(\log^2 n\log\log n\log k)$. With the result from \cite{fakcharoenphol2003tight}, it improves to $O(\log k \log^2n)$ as shown in the first entry in this table.}
    \label{table.generalized}
\end{table}

\noindent Generalized TSP is also referred to as Set TSP or One-of-a-Set TSP, or Group TSP in the literature, cf.~\cite{noon1993efficient,fischetti1997branch,guha1998approximation}.

\subsection{Clustered TSP}
\label{sec.clustered}
In the Clustered TSP, the nodes of $G$ are grouped into subsets. All nodes in $G$ must be traversed such that all nodes of one subset are traversed before the nodes in another subset. Usually, the subsets $P_1, \dots, P_m$ form a partition of the set of nodes $N$~\cite{laporte2002some,ding2007two}. Infrequently, this restriction is not made~\cite{guttmann2018approximation}. Sometimes, further restrictions on the subsets are made such as fixing a start and end node~\cite{laporte1997tabu}.

Most polynomial time approximability results for the Clustered TSP focus on the case when the clusters are disjoint. When there is additionally an order given in which the clusters should be traversed, a $5/3$ approximation is possible in polynomial time, whether the solution is to be a path or cycle or the start and end nodes are given or not. When the order of the clusters is not given, only less good approximation ratios are currently known. Several cases are covered in the literature depending on whether start and end nodes in each cluster are given or not. An overview of the polynomial time constant factor approximation results can be found in Table~\ref{table.clustered}.

\begin{table}[htp] \begin{tabular}{p{\textwidth}} $ $ \end{tabular}\vspace{-7ex}
\[
\begin{array}{lcrcr}
\textbf{Variant} & \textbf{Upper Bound} & \\ \hline\hline
\langle =1 \, | \,  =1\, | \,  \mbox{start; end; circuit; complete; undirected; partition(ordered)}\,|&\frac{5}{3}&\text{\cite{anily199953}} \unconfirmed\\
\hspace{5mm}c : E\mapsto \mathbb{R}_{>0},\mbox{(triangle)} \, | \, \min c(\sol)\rangle & &\\\hline
\langle =1 \, | \,  =1\, | \,  \mbox{start; end; complete; undirected; partition(ordered)}\,|&\frac{5}{3}&\text{\cite{anily199953}} \unconfirmed\\
\hspace{5mm}c : E\mapsto\mathbb{R}_{>0}, \mbox{(triangle)} \, | \, \min c(\sol)\rangle &&\\\hline

\langle =1 \, | \,  =1\, | \,  \mbox{circuit; complete; undirected; partition(ordered)}\,|&\frac{5}{3}&\text{\cite{anily199953}} \unconfirmed\\
\hspace{5mm}c : E\mapsto\mathbb{R}_{>0},\mbox{(triangle)} \, | \, \min c(\sol)\rangle & &\\\hline

\langle =1 \, | \,  =1\, | \,  \mbox{circuit; complete; undirected; partition(startend)}\, |&1.875&\text{\cite{kawasaki2020improving}} \unconfirmed\\
\hspace{5mm}c : E\mapsto \, \mathbb{R}_{\geq 0}, \mbox{(symmetric, triangle)} \, | \, \min c(\sol)\rangle &&\\\hline

\langle =1 \, | \,  =1\, | \,  \mbox{circuit; complete; undirected; partition(start)}\,|&\frac{5}{2}&\text{\cite{bao2017note}} \confirmed\\
\hspace{5mm}c : E\mapsto \, \mathbb{R}_{\geq 0}, \mbox{(triangle)} \, | \, \min c(\sol)\rangle &&\\\hline

\langle =1 \, | \,  =1\, | \,  \mbox{circuit; complete; undirected; partition(terminals)} \, |&1.714&\text{\cite{kawasaki2020improving}} \unconfirmed\\
\hspace{5mm}c : E\mapsto \, \mathbb{R}_{\geq 0}, \mbox{(triangle)} \, | \, \min c(\sol)\rangle &&\\\hline

\langle =1 \, | \,  =1\, | \,  \mbox{circuit; complete; undirected; partition}\,|&1.9&\text{\cite{bao2017note}} \confirmed\\
\hspace{5mm}c : E\mapsto\, \mathbb{R}_{\geq 0}, \mbox{(triangle)} \, | \,\min c(\sol)\rangle &&\\\hline

\langle =1 \, | \,  =1\, | \,  \mbox{complete; undirected; cover}\,|&4&\text{\cite{guttmann2018approximation}} \confirmed\\
\hspace{5mm}c : E\mapsto\, \mathbb{R}_{\geq 0}, \mbox{(triangle)} \, | \, \min c(\sol)\rangle &&\\ \hline \hline
\end{array}
\]
    \caption{Approximability results for the Clustered TSP.}
    \label{table.clustered}
\end{table}

Hybrid variants of the generalized and the clustered TSP are studied as well, \eg\cite{cosma2023hybrid,baniasadi2020transformation}.

\subsection{Traveling Purchaser Problem}
\label{sec.travelingPurchaser}

In the Traveling Purchaser Problem (TPP), a set of $m$ products is to be purchased at some nodes, which usually represent cities. Cities may offer the products at different prices and only a limited number of products may be available at each city. For each product, a demand $d_i$ is specified. For each city, the product availability specifies how many units of this product are available and the price for a unit of this product in this city is given. The aim is for the traveler to visit a subset of the cities and purchase at least the required demand of each product while minimizing the sum of purchasing and traveling costs. At each visited node $v$, a certain amount $\text{share}_i(v)$ of the available product is purchased. The sum of all $\text{share}_i(v)$ along the solution walk is then denoted $\text{share}_i(\sol_v)$. A good overview on the Traveling Purchaser Problem and its variants can be found in \cite{manerba2017traveling}, from which we formalized the basic definition of the problem using \tico in Table~\ref{table.traveling_purchaser}. The same definition of the problem is for example discussed in \cite{bontoux2008ant} and \cite{laporte2003branch}. Other variants use a modified range for the cost functions of demand and availability, e.g., in \cite{burstall1966heuristic}, the availability is either 0 or $\infty$. Alternatively as in \cite{batista2013traveling}, the demand is either 0 or 1, \ie the demand is not further specified.

\begin{table}[htp] \begin{tabular}{p{\textwidth}} $ $ \end{tabular}\vspace{-7ex}
\[
\begin{array}{lr}
\textbf{Variant} & \textbf{Reference} \\
\hline\hline

\langle =1 \, | \,  \leq 1; \mbox{always} \, | \,  \mbox{complete; directed} \, |  & \text{\cite{burstall1966heuristic}}  \unconfirmed\\
\hspace{5mm} c: E\mapsto\, \mathbb{R}_{> 0}; \{\text{avail}_i\}_{i=1}^m : V\mapsto\{0, \infty\}, \mbox{partial;}\, \{\text{price}_{i}\}_{i=1}^m : V\mapsto\mathbb{R}_{>0}\, | & \\
\hspace{5mm} \forall i \in \{1, \dots, m\} \,\,\,  \text{share}_i(\sol_V)  \geq d_i;\, 
\forall i \in \{1, \dots, m\} \,  \forall v \in V \, \text{share}_i(v) \leq  \text{avail}_i(v)  \, ; &\\
\hspace{5mm}\min c(\sol) + \sum_{i=1}^m\sum_{v \in \sol_V} \text{price}_i(v) \cdot \text{share}_i(v) \rangle \,&  \\ \hline

\langle =1 \, | \,  \leq 1; \mbox{always}\, | \,  \mbox{start; circuit; complete; directed} \, |  & \text{\cite{bontoux2008ant}} \confirmed\\
\hspace{5mm} c: E\mapsto\mathbb{R}_{>0}; \{\text{avail}_i\}_{i=1}^m : V\mapsto \mathbb{Z}_{\geq 0}, \mbox{partial;} \, \{\text{price}_{i}\}_{i=1}^m : V\mapsto\mathbb{R}_{\geq 0}\, |&\\ 
\hspace{5mm}\forall i \in \{1, \dots, m\} \,\,  \text{share}_i(\sol_V) \geq d_i;  \forall i \in \{1, \dots, m\} \,  \forall v \in V \, \text{share}_i(v) \leq  \text{avail}_i(v)  \, ; &\\
\hspace{5mm} \min c(\sol) + \sum_{i=1}^m\sum_{v \in \sol_V} \text{price}_i(v) \cdot \text{share}_i(v)\rangle \, &\\\hline

\langle =1 \, | \,  \leq 1; \mbox{always} \, | \,  \mbox{start; circuit; complete; directed} \, |  & \text{\cite{manerba2017traveling}} \confirmed\\
\hspace{5mm} c: E\mapsto\, \mathbb{R}_{\geq 0}; \{\text{avail}_i\}_{i=1}^m : V\mapsto\mathbb{Z}_{>0}, \mbox{partial;}\, \{\text{price}_{i}\}_{i=1}^m : V\mapsto\mathbb{R}_{>0}\, | & \\
\hspace{5mm} \forall i \in \{1, \dots, m\} \,\,\,  \text{share}_i(\sol_V)  \geq d_i;\, 
\forall i \in \{1, \dots, m\} \,  \forall v \in V \, \text{share}_i(v) \leq  \text{avail}_i(v)  \, ; &\\
\hspace{5mm}\min c(\sol) + \sum_{i=1}^m\sum_{v \in \sol_V} \text{price}_i(v) \cdot \text{share}_i(v) \rangle \,&  \\ \hline

\langle =1 \, | \,  \leq 1; \mbox{always} \, | \,  \mbox{start; circuit; undirected} \, |  & \text{\cite{laporte2003branch}} \confirmed\\
\hspace{5mm} c: E\mapsto\, \mathbb{R}_{\geq 0}; \{\text{avail}_i\}_{i=1}^m : V\mapsto\mathbb{Z}_{\geq 0}, \mbox{partial;}\, \{\text{price}_{i}\}_{i=1}^m : V\mapsto\mathbb{R}_{\geq0}\, | & \\
\hspace{5mm} \forall i \in \{1, \dots, m\} \,\,\,  \text{share}_i(\sol_V)  \geq d_i;\, 
\forall i \in \{1, \dots, m\} \,  \forall v \in V \, \text{share}_i(v) \leq  \text{avail}_i(v)  \, ; &\\
\hspace{5mm}\min c(\sol) + \sum_{i=1}^m\sum_{v \in \sol_V} \text{price}_i(v) \cdot \text{share}_i(v) \rangle \,&  \\ \hline \hline
\end{array}
\]
\caption{Variants of the Traveling Purchaser Problem. The cost function $c$ denotes the edge costs, the cost function $\text{avail}_i$ gives the availability of product $i$ at each node, and the cost function $\text{price}_i$ gives the cost of product $i$ at each node.}
    \label{table.traveling_purchaser}
\end{table}

The Traveling Purchaser Problem is polynomial-time solvable when the number of products $k$ is $O(\log n)$, where $n$ denotes the number of cities, or when the number of cities $n$ is $O(\log k)$~\cite{xiao2020parameterized}. Due to the hardness of the general problem~\cite{manerba2017traveling}, few algorithms with performance guarantees are known. To our knowledge, there exists only a single polynomial time approximation algorithm for the Traveling Purchaser Problem with a performance guarantee, which we present in Table~\ref{table.tpp}.
Note that \cite{ravi1999approximation} also presents a bi-criteria approximability result for a special case of TPP in the context of network design where all nodes in the graph must be added to the tour under an access cost budget. Fixed-parameter tractable algorithms and complexity results for the Traveling Purchaser Problem are given in~\cite{xiao2020parameterized}.
\begin{table}[htp] \begin{tabular}{p{\textwidth}} $ $ \end{tabular}\vspace{-7ex}
\[
\begin{array}{llr}
\textbf{Variant} & \textbf{Upper Bound} & \\
\hline\hline
\langle =1 \, | \,  \leq 1; \text{always};\, | \,  \mbox{start; circuit; complete; undirected}\,| & \max\big\{(1+\epsilon), &\\
\hspace{5mm} c: E\mapsto\mathbb{R}_{\geq 0},\mbox{metric; } &  \,\,\,\left(1+1/\epsilon\right)O(\log^3n\log\log n)\big\}  &\text{\cite{ravi1999approximation}} \\
\hspace{5mm}\{\text{avail}_i\}_{i=1}^m : V\mapsto\{0,1\}; \{\text{price}_{i}\}_{i=1}^m : V\mapsto\mathbb{R}_{\geq 0}\,|\\
\hspace{5mm} \forall i \in \{1, \dots, m\} \,\,\, \text{share}_i(\sol_V) \geq 1;   && \\ 
\hspace{5mm}\forall i \in \{1, \dots, m\} \,  \forall v \in V \, \text{share}_i(v) \leq  \text{avail}_i(v)  \, ; &\\
\hspace{5mm} \min c(\sol) + \sum_{i=1}^m\sum_{v \in \sol_V} \text{price}_i(v) \cdot \text{share}_i(v) \rangle \,&& \\ \hline \hline
\end{array}
\]
    \caption{Approximability result for the Traveling Purchaser Problem.}
    \label{table.tpp}
\end{table}

\subsection{The Profitable Tour Problem}
\label{sec.profitable-tour}

In the Profitable Tour Problem,\footnote{In the recent literature, the Profitable Tour Problem has also been frequently called Prize-Collecting TSP, which we use for the more general version presented in Sec.~\ref{sec.prizeCollecting} that furthermore includes a prize for each visited city and a lower bound on the total prizes that are to be collected.} the traveler receives a penalty when not visiting a city. The objective is to find the cheapest tour, when considering travel costs $c$ and penalties $p$.\footnote{The profitable tour problem considered in \cite{dell1995prize} has as objective to minimize the edge costs of the tour minus the sum of all profits along the tour.} 
An overview of approximability results is given in Table~\ref{table.TSPprofitabletour}. For complete graphs, the Profitable Tour Problem is at least as hard to approximate as the corresponding plain TSP variant, \eg by setting the penalties to twice the maximum edge cost. In the metric case, \ie for a given complete undirected graph with non-negative  edge costs satisfying the triangle inequality, Goemans and Williamson~\cite{goemans1995general} have presented a framework that yields a 2-approximation. This was later slightly improved to roughly 1.98 by~\cite{archer2011improved} and recently even more so as shown in Table~\ref{table.TSPprofitabletour}. The \emph{rooted} and \emph{unrooted} versions of the problem, \ie whether \emph{start} is true or false in our notation, are reducible to each other, while preserving approximation ratios~\cite{archer2011improved}. This is achieved by calling an algorithm for the unrooted problem with a sufficiently high penalty for the designated root or repeatedly calling one for the rooted problem with an uncovered node as root until each node has been covered by a tour. Thus, we specify only one of both versions in Table~\ref{table.TSPprofitabletour} based on the citation. 

\begin{table}[htp] \begin{tabular}{p{\textwidth}} $ $ \end{tabular}\vspace{-7ex}
\[
\begin{array}{lcr}
\textbf{Variant}  & \textbf{Upper Bound} & \\ \hline \hline

\langle =1 \,|\, \leq 1; \mbox{always}\,|\, \text{circuit; complete; directed}\,|\,   & O(1) & \text{\cite{nguyen2012approximating}} \confirmed  \text{\cite{TV2020}} \unconfirmed\\ 
\hspace{5mm}c :  E \mapsto \mathbb{R}, (\text{triangle}); p: V \mapsto \mathbb{R}_{\geq 0} \,|\, \min c(\sol) + \compl{p}(\sol)\rangle && \\ \hline

\langle =1 \,|\, \leq 1; \mbox{always}\,|\, \text{start; circuit; complete; undirected}\,|\,  & 1.599 &  \text{\cite{BlauthKN2024}} \confirmed \\ 
\hspace{5mm}c :  E \mapsto \mathbb{R}_{\geq 0}, (\text{triangle}); p: V \mapsto \mathbb{R}_{\geq 0} \,|\, \min c(\sol) + \compl{p}(\sol) \rangle & \\ \hline

\langle =1 \,|\, \leq 1; \mbox{always}\,|\, \text{start; complete; undirected}\,|\, & \frac{241}{141} & \text{\cite{archer2011improved}} \unconfirmed \\ 
\hspace{5mm} c :  E \mapsto \mathbb{R}_{\geq 0}, (\text{triangle}); p: V \mapsto \mathbb{R}_{\geq 0} \,|\, \min c(\sol) + \compl{p}(\sol) \rangle &&\\ \hline

\langle =1 \,|\, \geq 0; \leq\mbox{once}\,|\, \text{circuit; planar; undirected}\,|\,  & 1+\epsilon &  \text{\cite{BateniCEHKM2011}} \confirmed\\ 
\hspace{5mm}c :  E \mapsto \mathbb{R}_{\geq 0}; p: V \mapsto \mathbb{R}_{\geq 0} \,|\, \min c(\sol) + \compl{p}(\sol) \rangle & \\ \hline
\end{array}
\]
    \caption{Approximability results for the Profitable Tour Problem. 
     Note that the original bound obtained in~\cite{nguyen2012approximating} was $1+ \lceil \log_2(n) \rceil$. With the $O(1)$-approximation for ATSP in~\cite{TV2020}, it improves to the bound shown in the first entry in this table. Note that $\leq 1, \mbox{always}$ in the $\beta$ field can be replaced by $\geq 0, \mbox{once}$ in all variants with a cost function satisfying the triangle inequality, analogously to Fact~\ref{fact}.}
    \label{table.TSPprofitabletour}
\end{table}

\subsection{Quota TSP}
\label{sec.TSPWithProfits}

In the Quota TSP, each node is associated with a profit $q$. The aim is for the traveler to find a tour which visits each node at most once to collect the corresponding profit so that the total profit exceeds a given \emph{quota} $b$ while minimizing the travel cost.  Known approximability results are given in Table~\ref{table.TSPwithProfits}.

\begin{table}[htp] \begin{tabular}{p{\textwidth}} $ $ \end{tabular}\vspace{-7ex}
\[
\begin{array}{lcr}
\textbf{Variant} & \textbf{Upper Bound} & \\
\hline \hline
\langle =1 \,|\, \leq 1; \mbox{always} \,|\, \text{start; circuit; complete; undirected} \,|\, &5& \text{\cite{AusielloBLM18}} \confirmed\\
\hspace{5mm} c :  E \mapsto \mathbb{R}_{>0}, \text{(triangle)};q: V \mapsto \mathbb{Z}_{\geq 0} \,|\,\, q(\sol) \geq b; \min c(\sol)\rangle\\ \hline 

\langle =1 \,|\, \leq 1; \mbox{always} \,|\, \text{circuit; complete; undirected} \,|\,c :  E \mapsto \mathbb{R}_{>0}, \text{(triangle)} \,| &2&\text{\cite{garg2005saving}} \confirmed\\
\hspace{5mm} |V_\sol| \geq b; \min c(\sol)\rangle\\ \hline \hline
\end{array}
\]
    \caption{Approximability results for the Quota TSP. Note that $\leq 1, \mbox{always}$ in the $\beta$ field can be replaced by $\geq 0, \mbox{once}$ in all variants with a cost function satisfying the triangle inequality, analogously to Fact~\ref{fact}.}
    \label{table.TSPwithProfits}
\end{table}

In the literature, the Quota TSP is sometimes also referred to as TSP with Profits. When the task is to visit at least $k$ cities at minimum travel costs, this problem is also called $k$-TSP. However, the usage of names varies across publications. For example in~\cite{feillet2005traveling}, TSP with Profits is used as an umbrella term for the Profitable Tour problem, the Orienteering problem, and the Prize-Collecting TSP.

\subsection{Prize-collecting TSP}
\label{sec.prizeCollecting}
The Prize-collecting TSP is a combination of the Profitable Tour Problem (Section~\ref{sec.profitable-tour}) and the Quota TSP (Section~\ref{sec.TSPWithProfits}). As in the Profitable Tour Problem, the traveler receives a penalty $p$ when not visiting a city. Furthermore, as in the Quota TSP, each node may be associated with a profit $q$~\cite{bienstock1993note}, which the traveler receives if the node is visited. The aim of the traveler is to minimize the sum of all edge costs and penalties along the tour, while collecting a total profit, which is the sum of profits of visited nodes, of at least $b$ - where $b$ is also called the \emph{quota}~\cite{AusielloBLM18}. If $b=0$, the Prize-collecting TSP reduces to the Profitable Tour Problem. If $p(v)=0$ for all nodes $v$, the Prize-collecting TSP reduces to the Quota TSP. Table~\ref{table.pricecollectingVariants} shows these variants in entries (a) and (b), resp.

\begin{table}[htp] \begin{tabular}{p{\textwidth}} $ $ \end{tabular}\vspace{-7ex}
\[
\begin{array}{clr}
& \textbf{Variant} & \textbf{Reference} \\ \hline \hline
 & \langle =1 \, | \, \leq 1; \mbox{always} \, | \, \mbox{circuit; complete; directed} \, | \, & \text{\cite{marques2019quota}} \confirmed \\
 & c: E \mapsto \mathbb{R}; p: V \mapsto \mathbb{R}_{\geq 0}  \, | \, \min c(\sol) + \compl{p}(\sol) \rangle \\ \hline
 
(a) & \langle =1 \, | \, \leq 1; \mbox{always} \, | \, \mbox{circuit; complete; undirected} \, | \, & \text{\cite{marques2019quota}}\confirmed  \\
 & c: E \mapsto \mathbb{R}; p: V \mapsto \mathbb{R}_{\geq 0}  \, | \, \min c(\sol) + \compl{p}(\sol) \rangle \\ \hline

(b) & \langle  =1 \, | \, \leq 1; \mbox{always} \, | \, \mbox{circuit; complete; undirected} \, | \, & \text{\cite{berube2009branch}}  \unconfirmed\\
& c: E \mapsto \mathbb{R}_{\geq 0} \mbox{, (triangle)}; \, q: V \mapsto \mathbb{R}_{\geq 0}  \, | \,   q(\sol) \geq b \, ; \, \min c(\sol) \, \rangle & \\ \hline

(c) & \langle  =1 \, | \, \leq 1; \mbox{always} \, | \, \mbox{circuit; complete; undirected} \, | \, & \text{\cite{balas1989prize}} \unconfirmed\\
& c: E \mapsto \mathbb{R}; \, q: V \mapsto \mathbb{R}_{\geq 0} \, ; \, p: V \mapsto \mathbb{R}_{\geq 0} \,| \,  q(\sol) \geq  b \, ; \,
\min c(\sol)+ \compl{p}(\sol) \, \rangle \\ \hline \hline

\end{array}
\]
    \caption{Variants of the Prize-Collecting TSP.}
    \label{table.pricecollectingVariants}
\end{table}

In the case of undirected graphs, the following result holds: Given an $\alpha$-approximation for (a) and a $\beta$-approximation for (b), we obtain an $(\alpha + \beta)$-approximation for variant (c) in Table~\ref{table.pricecollectingVariants} by \emph{concatenating} both solutions, \ie the traveler first visits all cities according to the tour for the Quota TSP, and before returning to the starting point, the traveler visits all cities of the Profitable Tour that have not been visited yet~\cite{AusielloBLM18}. This yields a constant factor approximation for the Prize-collecting TSP using the known constant factor approximations from the previous two subsections.

Similarly to the name TSP with profits, the name Prize-collecting TSP is also often used as an umbrella term to denote all variants of the TSP where the traveler collects some profit or a prize when visiting a node such as the Quota TSP, the Profitable Tour Problem, and the Orienteering Problem, see for example~\cite{balas1989prize}. Even if all penalties are equal to zero this name is frequently used. 

\subsection{Orienteering Problem}
\label{sec.orienteering}

In the Orienteering Problem~\cite{golden1987orienteering,vansteenwegen2011orienteering}, the traveler maximizes the profit $q$ without exceeding a given budget $b$ for the total travel cost.  An $\alpha$-approximation algorithm for the Orienteering Problem with unit profits,  \ie all node costs are equal to 1, yields an $\alpha \cdot(1 + o(1))$-approximation algorithm for the Orienteering Problem with profits in $\mathbb{R}_{\geq 0}$~\cite[Lemma~2.6]{korula2010approximation}. The Orienteering Problem occurs in the literature in various variants as shown in Table~\ref{table.orienteeringVariants}.
\begin{table}[htp] \begin{tabular}{p{\textwidth}} $ $ \end{tabular}\vspace{-7ex}
\[
\begin{array}{lr}
\textbf{Variant} & \textbf{Reference} \\ \hline \hline

\langle  =1 \, | \, \leq 1; \mbox{always} \, | \, \mbox{start; end; complete; undirected} \, | \, & \text{\cite{vansteenwegen2011orienteering}} \confirmed\\
\hspace{5mm} c: E \mapsto \mathbb{R}; \, q: V \mapsto \mathbb{R}_{\geq 0} \, | \,  c(\sol) \leq b; \max q(\sol) \rangle \, \\ \hline

\langle  =1 \, |\, \leq 1; \mbox{always} \, | \, \mbox{start; end; undirected} \, | \, &\text{\cite{kara2016new}} \unconfirmed\\
\hspace{5mm} c: E \mapsto \mathbb{R}; q: V \mapsto \mathbb{R}\, | \, c(\sol) \leq b; \max q(\sol) \rangle \, &   \\ \hline %

\langle =1 \, | \, \leq 1; \mbox{always} \, | \, \mbox{circuit; complete; undirected} \, | \, & \text{\cite{leifer1994strong}} \unconfirmed\\
\hspace{5mm} c: E \mapsto \mathbb{R}\mbox{, (triangle)}; \,  q: V \mapsto \mathbb{R}_{>0} \,| \, c(\sol) \leq b; \max q(\sol) \rangle\, \\ \hline

\langle =1 \, | \, \geq 0; \mbox{once} \, | \, \mbox{circuit; undirected} \, | \, & \text{\cite{kataoka1988algorithm}} \unconfirmed\\
\hspace{5mm} c: E \mapsto \mathbb{R}_{\geq 0}; \, q: V \mapsto \mathbb{R}_{\geq 0} \, | \, c(\sol) \leq b; \max q(\sol) \rangle \, & \\ \hline  

\langle =1 \, | \, \leq 1; \mbox{always} \, | \, \mbox{circuit; complete; undirected} \, | \, & \text{\cite{awerbuch1998new}} \confirmed\\
\hspace{5mm} c: E \mapsto \mathbb{R}_{\geq 0}; \, q: V \mapsto \mathbb{R}_{\geq 0}\, |
\, c(\sol) \leq b; \max q(\sol) \rangle \, & \\ \hline \hline

\end{array}
\]
    \caption{Variants of the Orienteering Problem.}
    \label{table.orienteeringVariants}
\end{table}
Approximability results for the Orienteering Problem are shown in Table~\ref{table.orienteering}. 

\begin{table}[htp] \begin{tabular}{p{\textwidth}} $ $ \end{tabular}\vspace{-7ex}
\[
\begin{array}{lcrcr}
\textbf{Variant} & \textbf{Lower Bound} & & \textbf{Upper Bound} & \\
\hline \hline

\langle =1 \,|\, \geq 0; \mbox{once} \,|\, \text{circuit; undirected} \,|&&& 2&\text{\cite{paul2017prize,paul2020budgeted}} \confirmed\\
\hspace{5mm}c :  E \mapsto \mathbb{R}_{\geq 0} \,|\, c(\sol) \leq b;\max |V_\sol| \rangle\\ \hline

\langle =1 \,|\, \geq 0; \mbox{once} \,|\, \text{start; end; undirected} \,| &&& 2+\epsilon& \text{\cite{chekuri2012improved}} \confirmed \\
\hspace{5mm}c :  E \mapsto \mathbb{R}_{\geq 0} \,|\, c(\sol) \leq b; \max |V_\sol| \rangle &&&\\ \hline

\langle =1 \,|\, \geq 0; \mbox{once} \,|\, \text{start; directed; circuit} \,| &&& 
\min \big\{ O(\log^2 OPT), & \text{\cite{chekuri2012improved}} \confirmed \\
\hspace{5mm}c :  E \mapsto \mathbb{R}_{\geq 0}, (\text{triangle}) \,|\, c(\sol) \leq b; \max |V_\sol| \rangle &&&
O \left( \log n /\log \log n \right) \big \}& \text{\cite{nagarajan2011directed}} \text{\cite{TV2020}} \unconfirmed\\ \hline

\langle =1 \,|\, \geq 0; \mbox{once}; \,|\, \text{start; undirected; complete} \,|  &\frac{1481}{1480}&\text{\cite{blum2007approximation}}&2+\epsilon& \text{\cite{chekuri2012improved}} \confirmed\\
\hspace{5mm}c: E \mapsto \{1,2\}, \text{metric} \,|\, c(\sol) \leq b; \max |V_\sol| \rangle\\ \hline  

\langle =1 \,|\, \leq 1; \mbox{always} \,|\, \text{start; undirected; complete} \,|&&& 1+\varepsilon&\text{\cite{ChenHar-Peled2008}} \unconfirmed \\
\hspace{5mm}c: E \mapsto \mathbb{R}_{\geq 0}, \text{euclidean fixed dim} \,|\, \\
\hspace{5mm}
c(\sol) \leq b; \max |V_\sol|\rangle\\ \hline \hline

\end{array}
\]
    \caption{Approximability results for the Orienteering Problem. $OPT \leq n$ is the number of nodes visited by an optimal solution.  Note that the original bound obtained in \cite{nagarajan2011directed} was $O(\log^2 n / \log \log n)$. With the $O(1)$-approximation for ATSP in~\cite{TV2020}, it improves to the bound shown in the third entry in this table. Note that $\leq 1, \mbox{always}$ in the $\beta$ field can be replaced by $\geq 0, \mbox{once}$ in the last entry, analogously to Fact~\ref{fact}.}
    \label{table.orienteering}
\end{table}

There is also a quasi-polynomial time $O(\log^2 k)$-approximation for the orienteering problem~\cite{chekuri2005recursive}. Papers introducing the Orienteering Problem frequently remark that this variant of the TSP is also called the Selective TSP \cite{kara2016new}, and vice versa \cite{laporte1990selective}. When taking a closer look at the definitions, one notices a subtle difference though. Most papers that use the name Selective TSP, usually assume that the traveler returns to the starting node, whereas most papers that use the term Orienteering Problem assume that a fixed start and end node are given, which are distinct from one another. The Selective TSP is also sometimes denoted as the Maximum Collection Problem, see for example~\cite{kataoka1988algorithm}. The literature also mentions Bank Robber Problem as another name for the Orienteering Problem~\cite{arkin1998resource,awerbuch1998new}. Historically, the Orienteering Problem was sometimes referred to as the Generalized TSP, see~\cite{golden1987orienteering}.

\subsection{Time-dependent TSP} 
\label{sec.time-dependent}
In the Time-dependent TSP, the edge costs or node costs change over time. For example in~\cite{gouveia1995classification,picard1978time,bigras2008time}, the costs of an edge or node usually depend on its position in the solution sequence. In~\cite{hansknecht2021dynamic}, the costs of an edge depend on the actual time at which the edge starts being traversed. In the moving-target or kinetic variant of the time-dependent TSP, targets move with a fixed constant speed in a fixed direction~\cite{hammar2002approximation, helvig2003}. A flexible timing scheme where traveling times depend on the time of the day is considered in~\cite{malandraki1996restricted}. A good overview over the timing models used in variants of time-dependent TSP and VRP problems is given in~\cite{gendreau2015time}.

\begin{table}[htp] \begin{tabular}{p{\textwidth}} $ $ \end{tabular}\vspace{-7ex}
\[
\begin{array}{lcrcr}
\textbf{Variant} & \textbf{Lower Bound} & & \textbf{Upper Bound} & \\ \hline \hline

\langle =1 \, | \,  =1\, | \,  \mbox{circuit; complete; undirected} \, | \,& 2^{\Omega(\sqrt{n})} & \text{\cite{hammar2002approximation}} \confirmed\\
\hspace{5mm}  c :  E \mapsto \mathbb{R}_{\geq 0}, \mbox{kinetic} \, | \, \min c(\sol) \rangle^{\oplus 1} &  &  & &\\ \hline

\langle =1 \, | \,  =1\, | \,  \mbox{circuit; complete; undirected} \, | \, & 2 & \text{\cite{hammar2002approximation}} \confirmed &   2+\epsilon  & \text{\cite{helvig1998moving,helvig2003}} \confirmed\\
\hspace{5mm}  c :  E \mapsto \mathbb{R}_{\geq 0},  \mbox{kinetic} \, | \, \min c(\sol) \rangle^{\oplus 2} &  &  & &\\ \hline

\langle =1 \, | \,  =1\, | \,  \mbox{circuit; complete; undirected} \, | \, &  &  &   1+\epsilon  & \text{\cite{hammar2002approximation}} \confirmed\\
\hspace{5mm}  c :  E \mapsto \mathbb{R}_{\geq 0},  \mbox{kinetic} \, | \, \min c(\sol) \rangle^{\oplus 3} &  &  & &\\ \hline

\langle =1 \, | \,  =1\, | \,  \mbox{circuit; complete; undirected} \, | \, 
&  & & 2-\frac{2}{3k} & \text{\cite{broden2004online}} \confirmed\\
\hspace{5mm} c :  E \mapsto \{1, 2\}, \mbox{poszone}(k)  \, | \, \min c(\sol)\rangle &  & & & \\ \hline

\langle =1 \, | \,  =1\, | \,  \mbox{circuit; complete; undirected} \, | \, & \omega(1)\footnotemark &  \text{\cite{broden2004online}} \confirmed&  &\\
\hspace{5mm} c : E \mapsto \mathbb{R}_{>0}, \mbox{euclidean}, \mbox{costzone(2)} \, | \, \min c(\sol)\rangle  &&&&\\ \hline

\langle =1 \, | \,  =1\, | \,  \mbox{circuit; complete; undirected} \, | \,  & & & \frac{5}{3} &  \text{\cite{broden2004online}} \confirmed\\
\hspace{5mm} c :  E \mapsto \{1,2\}, \mbox{costzone(2)} \, | \, \min c(\sol) \rangle &&&&\\ \hline

\langle =1 \, | \,  =1\, | \,  \mbox{start; circuit; complete; bidirected} \, | \,  & \infty & \text{\cite{hansknecht2021dynamic}} \confirmed& &  \\
\hspace{5mm} c :  E \mapsto \mathbb{R}_{\geq 0}, \mbox{(triangle)}, \mbox{time} \,| \, \min c(\sol) \rangle \, \\ \hline \hline
\end{array}
\]
\begin{description}[leftmargin=0.6cm,labelindent=0cm]
\item [$\oplus 1$:] 
    \begin{itemize}
    \item[--] An instance consists of $n$ moving points in the euclidean plane that are described by their respective starting points at time $0$ and their two-dimensional velocity vectors in polar coordinates.
    \item[--] The traveler starts at time $0$ at the origin of the coordinate system. The speed of the traveler is strictly larger than the speed of any moving target. It is often normalized to $1$.
    \end{itemize}
\item[$\oplus 2$:]  
    \begin{itemize}
    \item[--] The statements for $\oplus 1$ hold.
    \item[--] The lower bound for the approximability result holds when there are at least two moving targets or when the speed of each moving target is exactly half of the speed of the traveler.
    \item[--] The upper bound for the approximability result holds when there are at most  $O\Big(\frac{\log n}{\log \log n}\Big)$ moving targets. The other targets have speed 0 and are stationary.
     \end{itemize}
\item[$\oplus 3$:] The statements for $\oplus 1$ hold. All $n$ targets move with the same fixed constant speed in a fixed direction.
\end{description}
    \caption{Approximability results for the Time-dependent TSP. Recall that the values $\mbox{poszone}(k)$ and $\mbox{costzone}(k)$ of the attribute \emph{temporal} are defined in Subsection~\ref{section:costfield}.}
    \label{table.time-dependent}
\end{table}

\footnotetext{{In fact, the same argument as in~\cite{broden2004online} also holds for any polynomial-time computable function, thus, the entry could also be $\infty$}.}

When all $n$ targets and the traveler move on the same line  (restriction to one dimension, acyclic graph) and traveler and targets move with constant speed, an exact optimal algorithm of complexity $O(n^2)$ exists~\cite{helvig1998moving,helvig2003}. The inapproximability result for the metric Time-dependent TSP~\cite{hansknecht2021dynamic}, see last line of Table~\ref{table.time-dependent}, demonstrates that approximability results for the metric "standard" TSP do not carry over to the Time-dependent TSP.   Any $\alpha$-approximation of the asymmetric "standard "TSP yields an ($\alpha \lambda$)-approximation for the Time-dependent TSP when the time-dependent cost functions are of low variance, \ie bounded by a product of a static cost underestimator with a constant $\lambda \geq 1$~\cite{hansknecht2021dynamic}.

The Time-dependent Orienteering Problem requires the traveler to visit a maximum number of targets within a given deadline. Under the assumption that the ratio between the maximum and minimum traveling times between any two targets is constant, a $2 + \epsilon$-approximation algorithm exists~\cite{fomin2002approximation}. The kinetic variant of the Time-dependent TSP is also denoted as the Moving-Target TSP~\cite{helvig2003} or the Kinetic TSP~\cite{hammar2002approximation}. In the Moving-Target TSP with Resupply variant~\cite{helvig2003}, a single traveler must always return to the origin after visiting each target.

\subsection{TSP with Time Windows}
\label{sec.timewindows}

In the TSP with Time Windows, each node $v$ is associated with a time window $[r(v),d(v)]$ during which the node must be visited~\cite{christofides1981state,savelsbergh1985local}. Travelers may arrive at the node $v$ before the \emph{release time} $r(v)$, in which case they need to wait until $r(v)$ and usually incur a wait cost $w$ that extends the traveling time. Only seldomly, the waiting time is not increasing the costs of a tour, \eg~\cite{ascheuer2001solving}. In most variants, visiting the node $v$ after the \emph{deadline} $d(v)$ would not be considered a valid solution. Moreover, it is often permitted that some of the nodes may have infinite deadlines. It is commonly assumed that there is no preemption of a job when a traveler visits a node. This means, if a traveler visits a node, the job at this node is completed once and for all leading to node costs $h(v)$,  representing the \emph{handling time} at node $v$. The expression $c(\sol_{<i}) + w(\sol_{\leq i}) + h(\sol_{<i})$ thus denotes the arrival time of the traveler at node $i$ including the wait cost at node $i$. In case, the wait cost or handling time are 0, the expression reduces accordingly. \tico definitions for typical variants of the TSP with Time Windows are summarized in Table~\ref{table.TSPTimeWindowVariants}.

\begin{table}[htp] \begin{tabular}{p{\textwidth}} $ $ \end{tabular}\vspace{-7ex}
\[
\begin{array}{lr}
\textbf{Variant} & \textbf{Reference} \\
\hline \hline
\langle  =1 \, | \,  \geq 1;  \mbox{once}\, | \,  \mbox{circuit; undirected} \, | \, c: E \mapsto \mathbb{R}_{\geq 0}; \, w: E \mapsto \mathbb{R}_{\geq 0}, \mbox{waiting}; h: V \mapsto \mathbb{R}_{\geq 0} \, | \, & \text{\cite{christofides1981state}} \confirmed\\
\hspace{5mm} \forall \, i \in \{0, \dots, k\} : \, r(v_i) \leq c(\sol_{<i}) + w(\sol_{\leq i}) + h(\sol_{<i}) \leq d(v_i) - h(\sol_{i}); \min c(\sol)+w(\sol)+h(\sol) \rangle \, \\ \hline 

\langle  =1 \, | \,  =1;  \mbox{always}\, | \,  \mbox{circuit; complete; undirected} \, | \,  c: E \mapsto \mathbb{R}_{\geq 0}, \mbox{(triangle)}; \, w: E \mapsto \mathbb{R}_{\geq 0}, \mbox{waiting} \, | \, & \text{\cite{savelsbergh1985local}} \unconfirmed\\
\hspace{5mm} \forall \, i \in \{0, \dots, k\} : \, r(v_i) \leq c(\sol_{<i}) + w(\sol_{\leq i}) \leq d(v_i);  \min c(\sol) + w(\sol) \rangle \, \\ \hline 

\langle =1 \, | \,  =1;  \mbox{always} \, | \,  \mbox{start; end; complete; bidirected} \, | \, c: E \mapsto \mathbb{R}_{\geq 0}; w: E \mapsto \mathbb{R}_{\geq 0}, \mbox{waiting} \, | \, & \text{\cite{gendreau1998generalized}} \unconfirmed \\

\hspace{5mm} \forall \, i \in \{0, \dots, k\} : \, r(v_i) \leq c(\sol_{<i}) + w(\sol_{\leq i}) \leq d(v_i); \min c(\sol)  \rangle \\  \hline  

\langle  =1 \, | \,  \geq 1;  \mbox{once} \, | \, \mbox{circuit; directed} \, | \,c: E \mapsto \mathbb{R}_{\geq 0}, \mbox{(triangle)}; \, w: E \mapsto  \mathbb{R}_{\geq 0},\mbox{waiting}; h: V \mapsto \mathbb{R}_{\geq 0} \, | \, & \text{\cite{ascheuer2001solving}} \unconfirmed\\ 
\hspace{5mm}  \forall \, i \in \{0, \dots, k\} : \, r(v_i) \leq c(\sol_{<i}) + w(\sol_{ \leq i}) + h(\sol_{<i}) \leq d(v_i) - h(\sol_{i}) ; \min c(\sol) + h(\sol) \rangle \, \\ \hline \hline 
\end{array}
\]
    \caption{Variants of the TSP with Time Windows.}
    \label{table.TSPTimeWindowVariants}
\end{table}

Various approximability results exist for the TSP with Time Windows, which are summarized in different figures based on the structure of the underlying graph. Table~\ref{table.time-windows-line-rt} summarizes approximation results for variants where the graph is a straight line, only release times are considered, and no deadlines are allowed. Table~\ref{table.time-windows} summarizes results for variants over trees or general graphs. The \textit{shoreline metric} considered in~\cite{psaraftis1990routing} is defined as follows:

\begin{definition}[Shoreline Metric~\cite{psaraftis1990routing}]
Let $1 \leq v_i  \leq v_k \leq v_j \leq n$ be nodes in the graph. Let $c(v_i, v_j)$ denote the costs of the edge from node $v_i$ to node $v_j$ as defined in Section~\ref{section.tsp-intro}. The shoreline metric satisfies the following conditions:
\begin{multicols}{3}
\begin{enumerate}
    \item $c(v_i,v_i) = 0$
    \item $c(v_i,v_j) = c(v_j,v_i)$
    \item $c(v_i,v_j) \geq c(v_i,v_k)$
    \item $c(v_i,v_j) \geq c(v_k,v_j)$
    \item $c(v_i,v_j) \leq c(v_i,v_k) + c(v_k,v_j)$
\end{enumerate}
\end{multicols}
\end{definition}

\begin{table}[htp] \begin{tabular}{p{\textwidth}} $ $ \end{tabular}\vspace{-7ex}
\[
\begin{array}{lcr}
\textbf{Variant} &   \textbf{Upper Bound} & \\ \hline \hline

\langle  =1 \, | \,  =1; \mbox{always} \, | \,  \mbox{circuit; complete; undirected} \, | \,    & 2Z/(Z+L)  & \text{\cite{psaraftis1990routing}}\\
\hspace{5mm} c: E \mapsto \mathbb{R}_{\geq 0}, \mbox{shoreline}; \, w: E \mapsto \mathbb{R}_{\geq 0}, \mbox{waiting} \, | \, &&\\ 
\hspace{5mm} \forall \, i \in \{0, \dots, k\} : \, r(v_i) \leq c(\sol_{<i}) + w(\sol_{\leq i}); \min c(\sol)+w(\sol) \, \rangle &&\\ \hline 

\langle  =1 \, | \,  =1; \mbox{always} \, | \,  \mbox{start; end; complete; undirected} \, | \,    & \min \{ 2(L+Z)/3L,   & \text{\cite{psaraftis1990routing}}\\
\hspace{5mm} c: E \mapsto \mathbb{R}_{\geq 0}, \mbox{shoreline}; \, w: E \mapsto \mathbb{R}_{\geq 0}, \mbox{waiting} \, | \, & (4Z-L)/2Z\} &\\ 
\hspace{5mm} \forall \, i \in \{0, \dots, k\} : \, r(v_i) \leq c(\sol_{<i}) + w(\sol_{\leq i}); \min c(\sol)+w(\sol) \,\rangle &&\\ \hline 

\langle  =1 \, | \,  \geq 1; \mbox{once} \, | \,  \mbox{circuit; undirected; path} \, | \,   & \frac{3}{2} & \text{\cite{karuno1998,karuno2002better}}\\
\hspace{5mm} c: E \mapsto \mathbb{R}_{\geq 0}); \,  w: E \mapsto \mathbb{R}_{\geq 0}, \mbox{waiting} ;\,  h: V \mapsto \mathbb{R}_{\geq 0} \, | \, &&\\
\hspace{5mm} \forall \, i \in \{0, \dots, k\} : \, r(v_i) \leq c(\sol_{<i}) + w(\sol_{\leq i}) + h(\sol_{<i}); \\
\hspace{5mm} \min c(\sol)+w(\sol)+h(\sol) \, \rangle^{\oplus 1} &&\\ \hline 

\langle  =1 \, | \, \geq 1; \mbox{once}\, | \,  \mbox{circuit; undirected; path} \, | \, c: E \mapsto \mathbb{R}_{\geq 0} ; \, h: V \mapsto \mathbb{R}_{\geq 0} \, | \, & \frac{3}{2} & \text{\cite{karuno2002better}}\\
\hspace{5mm} \forall \, i \in \{0, \dots, k\} : \,  c(\sol_{<i}) + h(\sol_{<i}) \leq d(v_i) - h(\sol_{i});  &&\\
\hspace{5mm}  \min \max_i \max \{0, c(\sol_{<i}) + h(\sol_{<i}) - d(v_i) \} \, \rangle \,&& \\ \hline 

\langle  =1 \, | \, \geq 1; \mbox{once}\, | \,  \mbox{circuit; undirected; path} \, | \,  c: E \mapsto \mathbb{R}_{\geq 0} ; \,  | \,& O(n^2) & \text{\cite{chan1995single,young1999single}}\\
\hspace{5mm} \forall \, i \in \{0, \dots, k\} : \, r \leq  c(\sol_{<i})  \leq d(v_i); \min c(\sol) \,\rangle&& \\ \hline \hline 

\end{array}
\]
\begin{description}
\item[$\oplus 1$:]
The solution is a so-called simple schedule. The traveler moves from the first node on the line to the last node and back and only chooses to visit a node and do the job at this node on the first move or on the second. 
\end{description}
\caption{Approximability results for the TSP with Time Windows on lines. The parameters $Z$ and $L$ denote the length of the shoreline where $Z= \sum_{i=1}^{n-1} c(v_i,v_{i+1})$ and $L=c(v_1,v_n)$ is the distance between node $1$ and node $n$. Hence, the upper bounds from \cite{psaraftis1990routing} are in all cases bounded by $2$. The result in~\cite{karuno1998,karuno2002better} also holds when the maximum lateness at nodes is minimized.}
    \label{table.time-windows-line-rt}
\end{table}

\begin{table}[htp] \begin{tabular}{p{\textwidth}} $ $ \end{tabular}\vspace{-7ex}
\[
\begin{array}{lcrcr}
\textbf{Variant} & \textbf{Lower Bound} & & \textbf{Upper Bound} & \\ \hline \hline

\langle  =1 \, | \,  \geq 1; \mbox{once} \, | \,  \mbox{circuit; bidirected; tree} \, | \, & &&  1 + \alpha & \text{\cite{karuno1997}} \unconfirmed\\
\hspace{5mm} c: E \mapsto \mathbb{R}_{\geq 0} ;\,  w: E \mapsto \mathbb{R}_{\geq 0}, \mbox{waiting}; \,  \\
\hspace{5mm} h: V \mapsto \mathbb{R}_{\geq 0} \, | \, &&\\
\hspace{5mm} \forall \, i \in \{0, \dots, k\} : \, r(v_i) \leq c(\sol_{<i}) + w(\sol_{\leq i}) + h(\sol_{<i}); \\
\hspace{5mm} \min c(\sol)+w(\sol)+h(\sol) \, \rangle^{\oplus 2} &&\\ \hline 

\langle  =1 \, | \,   \geq 1;  \mbox{once} \, | \,   \mbox{start; undirected; tree(b)} \, |  &&& 1 + \epsilon & \text{\cite{augustine2004linear}} \unconfirmed  \\
\hspace{5mm} c: E \mapsto \mathbb{R}_{\geq 0}, \mbox{metric}; w: E \mapsto \mathbb{R}_{\geq 0}, \mbox{waiting}; \,   \\
\hspace{5mm} h: V \mapsto \mathbb{R}_{\geq 0} \, | \, &&\\
\hspace{5mm} \forall \, i \in \{0, \dots, k\} : \, r(v_i) \leq c(\sol_{<i}) + w(\sol_{\leq i}) + h(\sol_{<i}); \\
\hspace{5mm} \min c(\sol)+w(\sol)+h(\sol)  \, \rangle^{\oplus 3}  && \\ \hline 

\langle  =1 \, | \,  = 1; \mbox{always} \, | \,  \mbox{circuit;  complete; undirected;} \, | \, &&&  \frac{5}{2} & \text{\cite{nagamochi1997}} \unconfirmed\\
\hspace{5mm} c: E \mapsto \mathbb{R}_{\geq 0} (\mbox{triangle}); \,  w: E \mapsto \mathbb{R}_{\geq 0}, \mbox{waiting}; \,   \\
\hspace{5mm} h: V \mapsto \mathbb{R}_{\geq 0} \, | \, &&\\
 \hspace{5mm} \forall \, i \in \{0, \dots, k\} : r(v_i) \leq c(\sol_{<i}) + w(\sol_{\leq i}) + h(\sol_{<i});   \\
 \hspace{5mm} \min c(\sol)+w(\sol)+h(\sol) \, \rangle &&\\ \hline 
 
\langle =1 \, | \,  = 1; \mbox{always}  \, | \,  \mbox{circuit; complete; undirected} \, | \, & 2 - \epsilon & \text{\cite{bockenhauer2007}} \confirmed &  \frac{5}{2} &  \text{\cite{bockenhauer2007}} \confirmed\\
\hspace{5mm} c: E \mapsto \mathbb{R}_{\geq 0}, \mbox{metric} \, | \,  \\
\hspace{5mm} \forall \, i \in \{0, \dots, n\} : \, c(\sol_{<i}) \leq d(v_i); \min c(\sol) \, \rangle^{\oplus 4} &&&\\ \hline \hline
\end{array}
\]
    \begin{description}
\item[$\oplus 2$:] The traveler starts at the root of the tree. 
\item[$\oplus 3$:]  There is a constant number of leaves $b$ in the tree.
\item[$\oplus 4$:]  The number of nodes $v_i$ with a finite deadline $d(v_i)$ is bounded by a constant $k$.
\end{description}
\caption{Approximability and inapproximability result for the TSP with Time Windows on trees or general graphs. The value of $\alpha$ is instance-dependent and depends on the maximum release time $r_{max}$, the sum of all traveling times, and the sum of all handling times. It always satisfies $0 < \alpha \leq 1 $, \ie the bound is at most 2. The lower approximation bound in~\cite{bockenhauer2007} applies even if the number of deadlines is at most 2. Furthermore, it was shown that there is no fixed paramater-tractable approximation algorithm for this problem variant with approximation ratio $2-\epsilon$ for any $\epsilon > 0$ unless P = NP. According to the authors of~\cite{augustine2004linear}, their result also applies to other tour variants where either start and end nodes are defined or the traveler travels a circuit. The paper only considers the case where the start of the traveler is given.}
    \label{table.time-windows}
\end{table}

The TSP with Time Windows is also denoted as TSP with Time Constraints~\cite{christofides1981state, savelsbergh1985local}, Shore Line Problem~\cite{psaraftis1990routing}, Traveling Repairman Problem~\cite{tsitsiklis1992special, frederickson2012approximation}, Time-constrained TSP~\cite{karuno1997}, Time-Window TSP~\cite{gao2020approximation}, or Single VRP with release and handling times~\cite{karuno1997}. VRP variants with Time Windows occur in many practical applications, see for example~\cite{solomon1987algorithms,desrochers1988vehicle,desrochers1992new}. Approximability results for those VRP variants have for example been published in~\cite{khachay2018improved,guo2022online}.

\subsection{Orienteering Problem with Time Windows}
\label{sec.Orienteering-TimeWindow}

In the Orienteering Problem with Time Windows as many nodes as possible are visited within their time windows. The deadline $d(v)$ is usually a "soft" deadline where a profit $p$ is collected if and only if a node is visited before its deadline. Profits usually replace the handling times that are considered in the TSP with Time Windows. In some variants, deadlines can also be exceeded by a constant factor, see for example~\cite{bansal2004approximation,farbstein2019deadline}.

Tables~\ref{table.time-windows-maxReward3},~\ref{table.time-windows-maxReward} and~\ref{table.time-windows-maxReward2} summarize approximability results for the Orienteering Problem with Time Windows split into Table~\ref{table.time-windows-maxReward3} for general profits, Table~\ref{table.time-windows-maxReward} for only release times or only deadlines, and Table~\ref{table.time-windows-maxReward2} for release times and deadlines under unit profits. 

Further approximability results have also been obtained for the Orienteering Problem with Time Windows under a variable speed of the traveler, for example,~\cite{frederickson2012approximation,gao2020approximation}. Variable costs changing at most $k-1$ times considering $k$ time zones are studied in~\cite{broden2004online}. Quasi-polynomial time approximations have been published in~\cite{friggstad2017compact,friggstadSwamy2021,chekuri2005recursive}. The Orienteering Problem with Time Windows~\cite{bansal2004approximation,chekuri2004maximum} is also known as the Repairman Problem,  (Speeding) Delivery Man Problem~\cite{frederickson2012approximation}, 
TSP with Deadlines~\cite{bockenhauer2007,bockenhauer2009approximation}, Time Window Prize Collecting Problem~\cite{gao2020approximation}, Prize-Collecting TSP with Time Windows~\cite{bar2005approximating}, or Deadline TSP~\cite{tsitsiklis1992special,augustine2004linear,bansal2004approximation,bar2005approximating,chekuri2012improved,frederickson2012approximation,friggstadSwamy2021}.

\bigskip

\bigskip

\begin{table}[htp] \begin{tabular}{p{\textwidth}} $ $ \end{tabular}\vspace{-7ex}
\[
\begin{array}{lcr}
\textbf{Variant} &  \textbf{Upper Bound} & \\ \hline \hline

\langle  =1 \, | \,  \geq 0; \leq \mbox{once} \, | \,  \mbox{start; undirected} \, | \,&   3 \log_2^2 n &\text{\cite{bansal2004approximation}} \unconfirmed\\ 
\hspace{5mm} c: E \mapsto \mathbb{Z}_{\geq 0}; \,  p: V \mapsto \mathbb{Z}_{\geq 0} \, |&& \\ 
\hspace{5mm} \forall \, i \in \{0, \dots, k\}: r(v_i) \leq c(\sol_{<i}) \leq d(v_i); \max p(\sol) \, \rangle && \\ \hline

\langle  =1 \, | \,  \geq 0; \leq \mbox{once}; \, | \, \mbox{start; directed}  \, |  & (1 + \epsilon) 
(\lfloor \sigma \rfloor + 1) &\text{\cite{bar2005approximating}} \confirmed \\
\hspace{5mm} c: E \mapsto \mathbb{R}_{\geq 0}; \, h: V \mapsto \mathbb{R}_{\geq 0}; \, p: V \mapsto \mathbb{R}_{\geq 0} \, | &\\
\hspace{5mm} \forall \, i \in \{0, \dots, k\}:  r(v_i) \leq c(\sol_{<i}) + h(\sol_{<i}) \leq d(v_i) - h(v_i); \max p(\sol) \rangle \\ \hline \hline
\end{array}
\]
\caption{Approximability results for the Orienteering Problem with Time Windows with release times and deadlines under general profits. Note that the result in~\cite{bar2005approximating} also includes handling times. $L$ is the ratio in length between the largest and the smallest time window $L = L_{max}/L_{min}$. The density parameter $\sigma$ of an instance is defined as $\sigma = max_{u,v} \frac{|L_u|}{c(u,v) + c(v,u) + h(u) + h(v)}$ where $L_u$ is the length of the time interval of node $u$ and $c(i,j)$ is the distance from node $i$ to node $j$. $OPT \leq n$ is the number of nodes visited by an optimal solution.}
    \label{table.time-windows-maxReward3}
\end{table}

\begin{table}[htp] \begin{tabular}{p{\textwidth}} $ $ \end{tabular}\vspace{-7ex}
\[
\begin{array}{lcr}
\textbf{Variant}  &  \textbf{Upper Bound} & \\ \hline \hline

\langle  =1 \, | \,   \geq 0; \leq \mbox{once}\, | \,  \mbox{start; undirected} \, | \,&   3 \log_2 n & \text{\cite{bansal2004approximation}} \unconfirmed\\ 
\hspace{5mm} c: E \mapsto \mathbb{Z}_{\geq 0}; \, p: V \mapsto \mathbb{Z}_{> 0} \, | \, && \\ 
\hspace{5mm} \forall \, i \in \{0, \dots, k\}:  c(\sol_{<i}) \leq d(v_i); \max p(\sol) \, \rangle && \\ \hline

\langle  =1 \, | \,   \geq 0; \leq \mbox{once}\, | \,  \mbox{start; undirected} \, | \,&   O(\log D_{max}) & \text{\cite{bansal2004approximation}} \unconfirmed\\ 
\hspace{5mm} c: E \mapsto \mathbb{Z}_{\geq 0}; \, p: V \mapsto \mathbb{Z}_{> 0} \, | \, && \\ 
\hspace{5mm} \forall \, i \in \{0, \dots, k\}:  c(\sol_{<i}) \leq d(v_i); \max p(\sol) \, \rangle && \\ \hline

\langle  =1 \, | \,   \geq 0; \leq \mbox{once}\, | \,  \mbox{start; undirected} \, | \,&   3 \log_2 n & \text{\cite{bansal2004approximation}} \unconfirmed\\ 
\hspace{5mm} c: E \mapsto \mathbb{Z}_{\geq 0}; \, p: V \mapsto \mathbb{Z}_{> 0} \, | \, && \\ 
\hspace{5mm} \forall \, i \in \{0, \dots, k\}: r(v_i) \leq c(\sol_{<i}) ; \max p(\sol) \, \rangle && \\ \hline

\langle  =1 \, | \,   \geq 0; \leq \mbox{once}\, | \,  \mbox{start; undirected} \, | \,&   O( \log \frac{1}{\epsilon}) & \text{\cite{bansal2004approximation}} \unconfirmed\\ 
\hspace{5mm} c: E \mapsto \mathbb{Z}_{\geq 0}; \, p: V \mapsto \mathbb{Z}_{> 0} \, | \, && \\ 
\hspace{5mm} \forall \, i \in \{0, \dots, k\}:  c(\sol_{<i}) \leq (1 +\epsilon) d(v_i); \max p(\sol) \, \rangle && \\ \hline

\langle =1 \, | \,  \geq 0; \leq \mbox{once} \, | \,  \mbox{start; undirected} \, | \,  & \beta +1  &  \text{\cite{chekuri2004maximum}} \confirmed\\
\hspace{5mm} c: E \mapsto \mathbb{R}_{\geq 0}, \mbox{metric}\, | \,  \\
\hspace{5mm} \forall \, i \in \{0, \dots, k\}: c(\sol_{<i}) \leq d(v_i);   
 \max |V_\sol|  \rangle^{\oplus 2} &&\\ \hline

 \langle  =1 \, | \, \geq 0; \leq \mbox{once}\, | \,  \mbox{start; undirected; complete} \, | \, &   (1+\epsilon)  \alpha & \text{\cite{farbstein2019deadline}} \unconfirmed\\
\hspace{5mm} c: E \mapsto \mathbb{Z}_{\geq 0}, \mbox{metric}; p: V \mapsto \mathbb{Z}_{>0} \, | \, \\
\hspace{5mm} \forall \, i \in \{0, \dots, k\}: c(\sol_{<i}) \leq (1 +2\epsilon) d(v_i); \max p(\sol)  \, \rangle^{\oplus 1} && \\ \hline

\langle  =1 \, | \, \geq 0; \leq \mbox{once}\, | \,  \mbox{start; undirected; tree} \, | \, &     1+\epsilon& \text{\cite{farbstein2019deadline}} \unconfirmed\\ 
\hspace{5mm} c: E \mapsto \mathbb{Z}_{\geq 0}, \mbox{metric}; p: V \mapsto \mathbb{Z}_{>0} \, | \, \\
\hspace{5mm} \forall \, i \in \{0, \dots, k\}: c(\sol_{<i}) \leq (1 +\epsilon) d(v_i); \max p(\sol) \, \rangle^{\oplus 1} &  \\ \hline 

\langle  =1 \, | \, \geq 0; \leq \mbox{once}\, | \,  \mbox{start; undirected; tree} \, | \, &     1 & \text{\cite{farbstein2019deadline}} \unconfirmed\\ 
\hspace{5mm} c: E \mapsto \mathbb{Z}_{\geq 0}, \mbox{metric}; p: V \mapsto \mathbb{Z}_{>0} \, | \, \\
\hspace{5mm} \forall \, i \in \{0, \dots, k\}: c(\sol_{<i}) \leq (1 +\epsilon) d(v_i); \max p(\sol) \, \rangle^{\oplus 2} &  \\ \hline 

\langle =1 \, | \,  \geq 0; \leq \mbox{once};  \, | \,  \mbox{start; end;  directed} \, | \, & \min \big \{ O(\log^3  OPT), &\text{\cite{chekuri2012improved}} \confirmed \\
\hspace{5mm} c: E \mapsto \mathbb{Z}_{\geq 0}\, | \, &  O(\log^2 n/\log \log n) \big \} & \text{\cite{nagarajan2011directed}} \text{\cite{TV2020}} \unconfirmed\\
\hspace{5mm} \forall \, i \in \{0, \dots, k\} : \, c(\sol_{<i}) \leq d(v_i);  \max |V_\sol|  \rangle&&\\ \hline

\hline
\end{array}
\]
\begin{description}
\item[$\oplus 1$:] The number of distinct deadlines is a fixed constant independent of the input.
\item[$\oplus 2$:] The number of nodes with finite deadlines is a constant. 
\end{description}
\caption{Approximability results for the Orienteering Problem with Time Windows and either release times or deadlines only. The constant $\beta$ is the approximation ratio for the Orienteering Problem with time windows with a single deadline where the start and end of the tour are defined~\cite{bansal2004approximation}. The constant $\alpha$ is the approximation ratio for the prize-collecting TSP with time windows with at most $k$ deadlines~\cite{chekuri2004maximum} with $\alpha = 3$ in~\cite{farbstein2019deadline}. The constant $D_{max} = \max_{v \in V} d(v)$ is the maximum deadline in the graph. $OPT \leq n$ is the number of nodes visited by an optimal solution. Note that the bound in the last entry follows from ~\cite{chekuri2012improved}, the result in~\cite{nagarajan2011directed}, and the $O(1)$-approximation for ATSP in~\cite{TV2020}.}
\label{table.time-windows-maxReward}
\end{table}

\begin{table}[htp] \begin{tabular}{p{\textwidth}} $ $ \end{tabular}\vspace{-7ex}
\[
\begin{array}{lcl}
\textbf{Variant} &  \textbf{Upper Bound} & \\ \hline \hline

 \langle =1 \, | \,  \geq 0; \leq \mbox{once} \, | \,  \mbox{start; undirected} \, | \,  & \beta +1  &\text{\cite{chekuri2004maximum}} \confirmed\\
\hspace{5mm} c: E \mapsto \mathbb{R}_{\geq 0}, \mbox{metric} \, | \, \\
\hspace{5mm} \forall \, i \in \{0, \dots, k\}:  r(v_i) \leq c(\sol_{<i}) \leq d;   
 \max |V_\sol| \rangle^{\oplus 3} &&\\ \hline
 
\langle =1 \, | \, \geq 0; \leq \mbox{once}  \, | \,  \mbox{start; undirected} \, | \,  & \beta +2  &\text{\cite{chekuri2004maximum}} \confirmed\\
\hspace{5mm} c: E \mapsto \mathbb{R}_{> 0}, \mbox{metric} \, | \,\\
\hspace{5mm} \forall \, i \in \{0, \dots, k\}: r(v_i) \leq c(\sol_{<i}) \leq d(v_i);   
 \max |V_\sol| \rangle^{\oplus 4} &&\\ \hline

\langle  =1 \, | \,  \geq 0; \leq \mbox{once}; \, | \,   \mbox{start; undirected; path} \, |  &  4 + \epsilon &\text{\cite{bar2005approximating}} \confirmed \\
\hspace{5mm} c: E \mapsto \mathbb{R}_{\geq 0} \, | &\\
\hspace{5mm} \forall \, i \in \{0, \dots, k\}: r(v_i) \leq c(\sol_{<i}) \leq r(v_i)+1; 
 \max |V_\sol| \rangle\\ \hline

\langle  =1 \, | \,  \geq 0; \leq \mbox{once}; \, | \,   \mbox{start; undirected; path} \, |  &  (4 + \epsilon) \log_2 L &\text{\cite{bar2005approximating}} \confirmed\\
\hspace{5mm} c: E \mapsto \mathbb{R}_{\geq 0}\,  | &\\
\hspace{5mm} \forall \, i \in \{0, \dots, k\}: r(v_i) \leq c(\sol_{<i}) \leq d(v_i); 
 \max |V_\sol| \rangle\\ \hline
 
\langle  =1 \, | \,  \geq 0; \leq \mbox{once}; \, | \,  \mbox{start; directed} \, |  &  \lfloor \sigma \rfloor + 1 &\text{\cite{bar2005approximating}} \confirmed\\
\hspace{5mm} c: E \mapsto \mathbb{R}_{\geq 0} ;\, h: V \mapsto \mathbb{R}_{\geq 0} \, | &\\
\hspace{5mm} \forall \, i \in \{0, \dots, k\}: r(v_i) \leq c(\sol_{<i}) + h(\sol_{<i}) \leq d(v_i) - h(v_i); 
 \max |V_\sol| \rangle\\ \hline

\langle =1 \, | \,  \geq 0; \leq \mbox{once};  \, | \,  \mbox{start; end;  undirected} \, | \, & O(\max \{\log OPT, \log L\})  &\text{\cite{chekuri2012improved}} \confirmed \\
\hspace{5mm} c: E \mapsto \mathbb{Z}_{\geq 0}; w: E \mapsto \mathbb{Z}_{\geq 0}, \mbox{waiting} \, |&& \\
\hspace{5mm} \forall \, i \in \{0, \dots, k\} : \, r(v_i) \leq  c(\sol_{<i}) + w(\sol_{\leq i}) \leq d(v_i);  \max |V_\sol| \rangle&&\\ \hline

\langle =1 \, | \,  \geq 0; \leq \mbox{once};  \, | \,  \mbox{start; end; bidirected} \, | \, & O(\log^2 OPT  \cdot &\text{\cite{chekuri2012improved}} \confirmed \\
\hspace{5mm} c: E \mapsto \mathbb{Z}_{\geq 0}; w: E \mapsto \mathbb{Z}_{\geq 0}, \mbox{waiting}\, | &  \max \{\log OPT, \log L \})& \\
\hspace{5mm} \forall \, i \in \{0, \dots, k\} : \, r(v_i) \leq  c(\sol_{<i}) + w(\sol_{\leq i}) \leq d(v_i);  \max |V_\sol| \rangle&&\\ \hline

\langle =1 \,  | \, \geq 0; \leq \mbox{once} | \,  \mbox{start; end; undirected; tree} \, | \, &  3  &\text{\cite{frederickson2012approximation}} \\
 \hspace{5mm} c: E \mapsto \mathbb{R}_{\geq 0} \, | \, && \\
\hspace{5mm} \forall \, i \in \{0, \dots, k\}: r(v_i) \leq c(\sol_{<i}) \leq r(v_i)+1; \max |V_\sol| \, \rangle&& \\ \hline

\langle =1 \,  | \, \geq 0; \leq \mbox{once} | \,  \mbox{start; end; undirected} \, | \, &  6 +\epsilon  &\text{\cite{frederickson2012approximation}} \\
 \hspace{5mm} c: E \mapsto \mathbb{R}_{\geq 0} \, | \, && \\
\hspace{5mm} \forall \, i \in \{0, \dots, k\}: r(v_i) \leq c(\sol_{<i}) \leq r(v_i)+1; \max |V_\sol| \, \rangle && \\ \hline 

\langle =1 \,  | \, \geq 0; \leq \mbox{once} | \,  \mbox{start; end; undirected} \, | \, &  O(\log L)  &\text{\cite{frederickson2012approximation}} \\
 \hspace{5mm} c: E \mapsto \mathbb{R}_{\geq 0} \, | \, && \\
\hspace{5mm} \forall \, i \in \{0, \dots, k\}: r(v_i) \leq c(\sol_{<i}) \leq d(v_i); \max |V_\sol| \, \rangle^{\oplus 5} && \\ \hline 

\langle =1 \, | \, \geq 0; \leq \mbox{once}  \, | \,  \mbox{start; end; undirected} \, | \,  & 10 &\text{\cite{frederickson2012approximation}}\\
\hspace{5mm} c: E \mapsto \mathbb{R}_{> 0} \, |\\
\hspace{5mm} \forall \, i \in \{0, \dots, k\}: r(v_i) \leq c(\sol_{<i}) \leq d(v_i);   
 \max |V_\sol| \rangle^{\oplus 6} &&\\ \hline
\hline

\end{array}
\]
\begin{description}
\item[$\oplus 3$:] There is a single deadline.
\item[$\oplus 4$:] The number of distinct time windows is a fixed constant independent of the input.
\item[$\oplus 5$:] All start end and end times are integers.
\item[$\oplus 6$:] All time windows have length in $[1,2]$.
\end{description}
\caption{Approximability results for the Orienteering Problem with time windows considering release times and deadlines under unit profits. Note that one result in~\cite{bar2005approximating} also includes handling times. The constant $\beta$ is the approximation ratio for the Orienteering Problem with time windows with a single deadline where the start and end of the tour are defined~\cite{bansal2004approximation}. $L$ is the ratio in length between the largest and the smallest time window $L = L_{max}/L_{min}$. The density parameter of an instance is defined as $\sigma = \max_{u,v} \frac{|L_u|}{c(u,v) + c(v,u) + h(u) + h(v)}$ where $L_u$ is the length of the time interval of node $u$ and $c(i,j)$ is the distance from node $i$ to node $j$. $OPT \leq n$ is the number of nodes visited by an optimal solution.}
 \label{table.time-windows-maxReward2}
\end{table}

\FloatBarrier

\section{Extension of \tico to Variants of the Vehicle Routing Problem}
\label{section.vrp}

The \tico definition scheme can easily be extended to accommodate several travelers such that variants of the vehicle routing problem VRP can be defined.\footnote{See~\cite{toth2002vehicle} for an introduction and overview of the VRP.}
In particular, attributes and values from the classification schemes proposed in~\cite{bodingolden1981, desrochers1990} could be added to the respective \tico fields.  For example, the attributes \textit{vehicle scheduling} and \textit{route duration} originally belong to the \textit{vehicles}-field in~\cite{desrochers1990}, but would be added to the $\gamma$-field in \tico, because they characterize the tour of a vehicle. In order to illustrate a possible extension of \tico to VRP, we define a variant of the Capacitated Vehicle Routing Problem CVRP as described in the first publication introducing the CVRP~\cite{dantzig1959truck} in Figure~\ref{fig.cvrp}.

When defining problem variants with several travelers, one usually needs to refer to properties or costs relating to a single traveler rather than the entire collection of travelers. We denote the set of travelers by $T$ and if an attribute refers to a specific traveler rather than the set of travelers, we indicate this by adding an index $t$ to an attribute or cost function. In~\cite{dantzig1959truck}, the CVRP variant is defined over a complete graph with non-negative edge costs and each node is associated with a non-negative demand $d$. Edge costs are symmetric. Travelers (vehicles) deliver
items from a depot to the customers. Each vehicle must start at the depot, perform a circuit to visit each customer exactly once, and return to the depot. The paper first discusses a variant where all vehicles have the same capacity $cap$, but then also considers an extension to vehicles of different capacities $cap_t$. The sum of demands along each vehicle's circuit must not exceed the vehicle's capacity.

\begin{figure}[ht]
\fbox{
\parbox{\textwidth}{
\vspace*{3mm}
\begin{tabbing}
$\langle$ \= $\alpha:$ \= count $=|T|$; \\
                             \> \> capacity$_t$ = $c_t$;\\
\> $\beta:$                     \> traversals = 1;\\
                             \> \> visits = always;\\
\> $\gamma:$                    \> start = True; \\
                             \> \> complete = True; \\
                             \> \> directed = True; \\
                             \> \> $\forall \, t \in T$ circuit$_t$ = True\\
\> $\delta:$                    \> $c: E \mapsto \mathbb{R}_{\geq 0}$, symmetric;\\
                             \> \> $d: V \mapsto \mathbb{Z}_{> 0}$ ; \\
\> $\epsilon:$                 \>   $\forall \, t \in T: \,  d(\sol_{V_t}) \leq cap_t$; $ \min c(\sol) $ $\rangle$
\end{tabbing}

\bigskip
\begin{tabbing}
$\langle =|T|\,| =1; \, \text{always}|\,\text{start; circuit$_t$; complete; directed}\,|\,$ \\
\hspace{13mm} $c :  E \mapsto \mathbb{R}_{\geq 0},   \text{symmetric} ; d: V \mapsto \mathbb{Z}_{> 0} \,|\,  \forall \, t \in T: \,   d(\sol_{V_t}) \leq cap_t;  \min c(\sol) \rangle$ \\
\end{tabbing}
}}
\caption{\label{fig.cvrp}\tico definition of the Capacitated Vehicle Routing Problem CVRP described in~\cite{dantzig1959truck}.} 
\end{figure}

A recent overview of approximability results for variants of the CVRP is available on arxiv.org~\cite{chen2023survey}. Results were \eg obtained for the CRVP problem in the euclidean plane~\cite{haimovich1985bounds}, for a variant with unit demand, equal capacity of all vehicles and a fixed, but arbitrary number of depots under an edge cost function, which is determined by a fixed-dimensional euclidean metric~\cite{khachay2016ptas}, and for a variant with a linearly ordered set of time windows and equal capacity of all vehicles in the euclidean plane~\cite{khachay2019approximation}.  It is easy to see that \tico would facilitate the precise and compact definition of these variants using existing attributes and values and requiring only minimal extensions.

\section{Conclusion}
\label{section.conclusion}	
This paper provides the first systematic survey on the best currently known (in)approximability results for well-known TSP variants such as the "standard" TSP,  Path TSP, Bottleneck TSP, Maximum Scatter TSP,  Generalized TSP, Clustered TSP, Traveling Purchaser Problem, Profitable Tour Problem, Quota TSP, Prize-Collecting TSP, Orienteering Problem, Time-dependent TSP, TSP with Time Windows, and the Orienteering Problem with Time Windows. 
 
The foundation of this survey is the \tico definition scheme, which takes inspiration from a similar and successfully used definition scheme from the domain of scheduling problems. We propose \tico as a uniform, easy-to-use and extensible means for the formal and precise definition of TSP variants by introducing five fields: the $\alpha$-, $\beta$-, and $\gamma$-fields, which characterize the number of travelers, the targeted cities, and the tour, and the $\delta$- and $\epsilon$-fields, which define costs and possible objectives. For each field, a set of distinguished attributes and values is defined for the \tico 2025 version used in the survey.

Applying \tico to formally define the variant studied by a paper reveals subtle differences within the same named variant and also brings out the differences between the variants more clearly. This makes it easier to understand the approximability landscape and the assumptions under which certain results hold. Open gaps become more evident and results can be compared more easily. In summarizing the insights from our survey, we propose a \tico definition for each TSP variant. A brief discussion of how \tico can be used to define vehicle routing problems by adding new attributes and values to the definition scheme concludes the paper.

\medskip
\textbf{Acknowledgement:} We would like to thank the anonymous reviewers and the editor for their valuable feedback. We would like to thank the following authors for providing feedback on an earlier version of this paper: Xiaoguang Bao, Nili	Beck, Hans-Joachim Böckenbauer, Vincenzo Bonifaci, Chandra Chekuri, Dominique Feillet, Daniel Freund, Naveen Garg, Elizabeth Goldbarg, Imke Joormann, Michael Khachay, Philip Klein, Robert Kleinberg, Sascha Kurz, Michail Lampis, Daniele Manerba, Martin Nägele, Hung Nguyen, Bengt J. Nilsson, Harilaos N.\ Psaraftis, Jörg Rambau, Andr{\'a}s Seb{\H{o}, Shimon Shahar, Steven Skiena, Paolo Thoth, Vera Traub, Pieter Vansteenwegen, Santhosh S. Vempala, Jens Vygen, Benjamin Wah, Alex Zelikowsky.

\medskip

\textbf{Conflict of interest:}
Not applicable.

\bibliography{lit}

\begin{appendices}

\section{\tico 2025 at a Glance: EBNF and Set of Attributes and Values}
\label{appendixA}

\begin{figure}[H]
\fbox{
\parbox{0.8\linewidth}{
\begin{tabbing}
\tico \hspace{5mm} \=  $\Leftarrow$ \hspace{5mm} \=``$\langle \,$'' $\alpha$-\textit{field} \hspace{1mm} $\beta$-\textit{field}  
\hspace{1mm} $\gamma$-\textit{field} \hspace{1mm} $\delta$-\textit{field} \hspace{1mm} $\epsilon$-\textit{field} $\, ``\rangle$''\\
$\alpha$-\textit{field} \>  $\Leftarrow$ \>   (``$\alpha$'' $\mid$ ``traveler'')  ``:''   $\{$ \textit{attribute} ``;''  $\}$ \\
$\beta$-\textit{field} \>  $\Leftarrow$ \>   (``$\beta$'' $\mid$ ``target'')  ``:''   $\{$ \textit{attribute} ``;'' $\}$ \\
$\gamma$-\textit{field} \>  $\Leftarrow$ \>   (``$\gamma$'' $\mid$ ``tour'')  ``:''   $\{$ \textit{attribute} ``;''  $\}$ \\
$\delta$-\textit{field} \>  $\Leftarrow$ \>   (``$\delta$'' $\mid$ ``costs'')  ``:''   $\{$ \textit{cost function} ``;'' $\}$ \\
$\epsilon$-\textit{field} \>  $\Leftarrow$ \>   (``$\epsilon$'' $\mid$ ``objective'')  ``:''   
 $\{$\textit{objective} ``;''$\, \}$ \\
\textit{attribute} \> $\Leftarrow$\> \textit{name}  $\,$ \textit{value}\\
\textit{name}\> $\Leftarrow$ \> string\\
\textit{value}\> $\Leftarrow$ \> \textit{relation} $\,$ mathematical expression \\
\textit{relation} \> $\Leftarrow$ \> ``='' $\mid$ ``$\leq$'' $\mid$ ``$<$'' $\mid$ ``$\geq$'' $\mid$ ``$>$'' $\mid$ ``$\in$''\\
\textit{cost function} \>  $\Leftarrow$ \> \textit{name} ``:''  \textit{domain} ``$\mapsto$'' \textit{range}  $\,   \{$``,'' \textit{attribute}$\}$  \\
\textit{domain} \> $\Leftarrow$ \> set\\
\textit{range} \> $\Leftarrow$ \> set\\
\textit{objective} \> $\Leftarrow$ \> mathematical expression
\end{tabbing}
	}
}
	\caption{Recap of Figure~\ref{fig:ebnf-long}:  EBNF for \tico longhand notation. }
\end{figure}

\begin{figure}[H]
\fbox{
\parbox{0.8\linewidth}{
\begin{tabbing}
\tico \hspace{3mm} \= $\Leftarrow$ \hspace{1mm}  \= ``$\langle$'' \hspace{1mm}
$\alpha$-\textit{field} \hspace{1mm} ``$\mid$'' \hspace{1mm} 
$\beta$-\textit{field}  \hspace{1mm} ``$\mid$'' \hspace{1mm}
$\gamma$-\textit{field} \hspace{1mm} ``$\mid$'' \hspace{1mm} 
$\delta$-\textit{field} \hspace{1mm} ``$\mid$'' \hspace{1mm} 
$\epsilon$-\textit{field} \hspace{1mm} ``$\rangle$''\\
$\alpha$-\textit{field} \> $\Leftarrow$  \> $\{$\textit{attribute} ``;''$\}$ \textit{attribute} \\
$\beta$-\textit{field} \> $\Leftarrow$  \>  $\{$\textit{attribute} ``;''$\}$ \textit{attribute}  \\
$\gamma$-\textit{field} \> $\Leftarrow$  \> $\{$\textit{attribute} ``;''$\}$ \textit{attribute} \\
$\delta$-\textit{field} \>  $\Leftarrow$ \>  $\{$\textit{cost function} ``;'' $\}$ \textit{cost function}\\
$\epsilon$-\textit{field} \>  $\Leftarrow$ \> $\{$\textit{objective} ``;'' $\}$ \textit{objective}\\
\textit{attribute} \> $\Leftarrow$\> (\textit{name} $[$\textit{value}]) $\mid$ ($[$\textit{relation}] mathematical expression)\\
\textit{cost function} \>  $\Leftarrow$ \> \textit{name} ``:''  \textit{domain} ``$\mapsto$'' \textit{range}  $\,   \{$``,'' \textit{attribute}$\}$  
\end{tabbing}
	}
}
\caption{Recap of Figure~\ref{fig:ebnf-short}:  EBNF for \tico shorthand notation.}
\end{figure}

\begin{sidewaystable}
\begin{tabular}{|c|l|l|l|l|} \hline
Field       & Attribute & Type(s) & Values & Comments\\ \hline
$\alpha$    & count     & int | expression  &  $=1$ $\mid$ $=\Theta(1)$  $\mid$  $=k$ $\mid$  $ \geq 1 $ & $k=\Theta(1)$ is a fixed parameter\\ 
traveler    &           &                   &                        &  that is not part of the input\\ \hline
            & traversals & int $\mid$ expression $\mid$   & $=1$ $\mid$ $=d$ $\mid$ $=d(v)$ $\mid$ &  d - number of traversals \\ 
            &            &               & $\geq 1$ $\mid$ $\geq 1$ $\mid$ $\mathit{\geq 0}$ $\mid$ $\geq d(v)$ $\mid$ $\leq 1$ &  d(v) -  traversals depending on  nodes \\ \dotlinec
$\beta$     & visits    & string    & = always$\mid$ = once $\mid$ $\geq$ once $\mid$ $\leq$ once  &  \\ \dotlinec
 target     & group     & string    & partition(once) $\mid$cover(once)  $\mid$  & \\
            &           &           & partition($\geq$ once) $\mid$ cover($\geq$ once) &  \\ \dotlinec
            & covering  & string    & all(c, $\leq$ b) $\mid$  subset (c, $\leq$ b)  
                            & c - cost function, B - cost bound \\ \hline
            & start         & bool      & True $\mid$ False &     \\ \dotlinec
            & end           & bool      & True $\mid$ False &     \\ \dotlinec
            & circuit       & bool      & True $\mid$ False &     \\ \dotlinec
            & graphtype     & string    & complete $\mid$ strongly connected $\mid$ planar $\mid$ &     \\
$\gamma$             &               &           & path $\mid$ cycle $\mid$ binary tree $\mid$ & \\
tour              &               &           & tree(b)   & \\ \dotlinec
            & edgetype      & string    &  undirected $\mid$ directed $\mid$ bidirected &     \\ \dotlinec
            & precedences   & string    & atomic $\mid$ arbitrary  &     \\  \dotlinec
            & cluster       & string    & partition(\texttt{order}, \texttt{sequence}) $\mid$  & \texttt{order}=ordered\\
            &               &           & cover(\texttt{order}, \texttt{sequence})        & \texttt{sequence} = start $\mid$ startend  $\mid$ terminals\\\hline
            & domain        & string    &  E $\mid$ V &     \\ \dotlinec
            & range         & string    & any subset of $\mathbb{R}_{>0}$  &  \\ \dotlinec
$\delta$    &  property     &  string   & (\texttt{params}) $\mid$ &  \texttt{params} =  identity $\mid$triangle$\mid$ $\alpha$-triangle\\   
cost        &               &           & metric $\mid$ graphic $\mid$ planar $\mid$ subset planar $\mid$ euclidean $\mid$             &  symmetric  \\ 
function    &               &           &  euclidean fixed dim  $\mid$ euclidean plane $\mid$ grid(m,n) & \\ \dotlinec
            & partial       & bool      & True $\mid$False &     \\ \dotlinec
            & temporal      & string    & waiting $\mid$ costzone(k) $\mid$ poszone(k)  $\mid$ &   k - zones change at most k-1 times  \\ 
            &               &          &   time $\mid$ position $\mid$ kinetic    &  \\ \hline
 $\epsilon$  &  & string & free syntax &  example: $c_1(\sol) + c_21(\sol) \leq b$, $\min c_1(\sol)$    \\ 
objective  &  & &   &   \\ \hline
    \end{tabular}
\end{sidewaystable}

\clearpage

\section{Proposals for TSP Variant Definitions in \tico 2025 Longhand Notation}
\label{appendixB}

We propose a \tico longhand definition for each of the TSP variants that we discussed in Section~\ref{section.approximability}. The longhand definition subsumes the shorthand \tico definitions of the various variations. We use the \tico longhand notation with the following conventions:

\begin{itemize}
    \item All fields that must be defined for the problem variant are listed, for fields, which are absent, no values must be defined. These fields are greyed out.
    \item If a field can take on any of the possible values defined in Section~\ref{section.attributesvalues}, we use ``$\ast$''.
    \item If a field can take on several, but not all possible values, these values are separated by "or".
    \item For the range of cost functions, the most general value is given, which is often $\mathbb{R}$.
\end{itemize}

\vspace{10mm}

\begin{center}
\begin{sidewaystable}
\begin{tabular}{|c|c|c|c|c|c|} \hline
\textit{Field} & \textit{Attribute}  & "standard" & Path     & Bottleneck & Maximum Scatter \\ 
               &                     &  TSP       & TSP      & TSP        &  TSP            \\ \hline
$\alpha$       &  count              & $= 1$      & $= 1$    & $= 1$      & $= 1$\\ \hline
\multirow{3}{1em}{$\beta$}    
        & traversals  & $= 1$ or $\geq 1$ & $= 1$ or $\geq 1$ &  $= 1$ or $\geq 1$ & $= 1$ or $\geq 1$   \\ \dotlinea
        & visits      & \greyed           & \greyed           & \greyed    &  \greyed  \\ \dotlinea
        & group       & \greyed           & \greyed           & \greyed    &  \greyed \\ \dotlinea
        & covering    & \greyed           & \greyed           & \greyed    &  \greyed \\ \hline
\multirow{7}{1em}{$\gamma$}  
        & start       & false             & $*$              &  $*$     &  $*$  \\ \dotlinea
        & end         & false             & $*$             &  $*$     & $*$   \\ \dotlinea
        & circuit     & true              & false             & $*$      & $*$   \\ \dotlinea
        & graphtype   & $*$               & $*$               &  $*$       & $*$     \\ \dotlinea
        & edgetype    & $*$               & $*$               &  $*$       & $*$     \\ \dotlinea
        & precedences & \greyed           & \greyed           & \greyed    & \greyed \\ \dotlinea 
        & cluster     & \greyed           & \greyed           & \greyed    & \greyed \\ \hline
\multirow{4}{1em}{$\delta$}  
       &  domain/range & $c:  E \mapsto \mathbb{R}$ & $c:  E \mapsto \mathbb{R}$ & $c:  E \mapsto \mathbb{R}$        & c: $E \mapsto \mathbb{R}$  \\ \dotlinea
       & property      & $*$              & $*$              &   $*$         &  $*$   \\ \dotlinea
       & partial       & \greyed          & \greyed          & \greyed       & \greyed   \\ \dotlinea
       & temporal      & \greyed          & \greyed          & \greyed       &  \greyed  \\ \hline
$\epsilon$ & objective & $\min c(\sol)$ & $\min c(\sol)$ & $\min \max \{c(e) : e\in E_{\sol}\}$ &$\max \min \{c(e) : e\in E_{\sol}\}$   \\ \hline
\end{tabular}
\end{sidewaystable}

\end{center}

\begin{sidewaystable}
\begin{tabular}{|c|c|c|c|c|c|c|} \hline
\textit{Field}  & \textit{Attribute}  &  Generalized & Clustered & Traveling Purchaser & Time-Dependent & TSP with\\ 
                &                     &  TSP         & TSP       & Problem             &  TSP           &  Time Windows \\ \hline
$\alpha$        &  count             & $= 1$         &   $= 1$   & $= 1$               &  $= 1$               & $=1$ \\ \hline
\multirow{3}{1em}{$\beta$} 
    & traversals  &  $\leq 1$  & $= 1$   &  $\leq 1$  & $=1$       &  $=1$ or $\geq 1$  \\  \dotlineb
    & visits      &  \greyed            &  \greyed            &  always    & \greyed    & once \\ \dotlineb
    & group       &  $*$                &  \greyed            &  \greyed   & \greyed    & \greyed \\ \dotlineb
    & covering    &  \greyed            &  \greyed            &  \greyed   & \greyed    & \greyed \\ \hline
\multirow{7}{1em}{$\gamma$}  
        & start       &    $*$     & $*$      &  $*$        & $*$ & $*$ \\ \dotlineb
        & end         &    $*$     & $*$      &  $*$        & $*$ & $*$ \\ \dotlineb
        & circuit     &    $*$     & $*$      &  $*$        & $*$ & $*$ \\ \dotlineb
        & graphtype   &    $*$     & $*$      &  $*$        & $*$ & $*$ \\ \dotlineb
        & edgetype    &    $*$     & $*$      &  $*$        & $*$ & $*$ \\ \dotlineb
        & precedences &  \greyed   & \greyed  &   \greyed   & \greyed  & \greyed \\ \dotlineb
        & cluster     &  \greyed   &  $*$     &  \greyed    &  \greyed & \greyed \\ \hline
\multirow{6}{1em}{$\delta$}  
        & domain/range& $c:  E \mapsto \mathbb{R}$& $c:  E \mapsto \mathbb{R}$ & $c:  E \mapsto \mathbb{R}$  
                        &  $c:  E \mapsto \mathbb{R}$ & $c: E \mapsto \mathbb{R}_{\geq 0}$ \\ 
        & & & & $\{\text{avail}_i\}_{i=1}^m : V\mapsto\mathbb{R}_{>0}$ & & $w: E \mapsto \mathbb{R}_{\geq 0}$\\  
        & & & & $\{\text{price}_{i}\}_{i=1}^m : V\mapsto\mathbb{R}_{>0}$ & & $h: V \mapsto \mathbb{R}_{\geq 0}$\\  \dotlineb
        & property      & $*$           & $*$           & $*$       & $*$       & $*$  \\ \dotlineb
        & partial       & \greyed       & \greyed       & $*$       & \greyed   & \greyed  \\ \dotlineb
        & temporal      & \greyed       & \greyed       & \greyed   & time, position      & $w:$ waiting \\
        &               &               &                &          & costzone(k), poszone(k) & \\ 
        &               &               &                &          & kinetic & \\  \hline
$\epsilon$ 
        & objective     & $\min c(\sol) $  & $\min c(\sol) $ &   $\min c(\sol)$ & $\min c(\sol) $ &  $\min *$ ;\\
        & & & & & $+ \sum_{i=1}^m\sum_{v \in \sol_V} \text{price}_i(v) \cdot \text{share}_i(v)$                    
        &  $\forall \, i \in \{0, \dots, k\} : \, r(v_i) \leq $ \\
        & &&&&& $c(\sol_{<i}) + w(\sol_{\leq i}) + h(\sol_{<i}) \leq$ \\
        & &&&&& $d(v_i) - h(\sol_{i})$ \\  \hline
        
\end{tabular}
\end{sidewaystable}

\begin{sidewaystable}
\begin{tabular}{|c|c|c|c|c|c|c|} \hline
\textit{Field}  & \textit{Attribute} & Prize-collecting & Quota & Profitable Tour & Orienteering &  Orienteering Problem\\ 
                &                    & TSP              &  TSP &  Problem               &  Problem     & with Time Windows \\ \hline
$\mathbb{\alpha}$   
    & count             & $= 1$         & $= 1$         & $= 1$         & $= 1$     &  $= 1$    \\ \hline
\multirow{3}{1em}{$\beta$}   
    & traversals        & $\geq 0$      & $\geq 0$     & $\geq 0$    & $\geq 0$  & $\geq 0$  \\ \dotlineb
    & visits            & $\leq$ once   & $\leq$ once  & $\leq$ once  & $\leq$ once  & $\leq$ once    \\ \dotlineb
    & group             & \greyed       & \greyed       & \greyed       & \greyed    & \greyed    \\ \dotlineb
    & covering          & \greyed       & \greyed       & \greyed       & \greyed    & \greyed    \\ \hline
\multirow{7}{1em}{$\gamma$} 
    & start             & $*$           & $*$           & $*$           & $*$       & $*$   \\ \dotlineb
    & end               & $*$           & $*$           & $*$           & $*$       & $*$   \\ \dotlineb
    & circuit           & $*$           & $*$           & $*$           & $*$       & $*$   \\ \dotlineb
    & graphtype         & $*$           & $*$           & $*$           & $*$       & $*$   \\ \dotlineb
    & edgetype          & $*$           & $*$           & $*$           & $*$       & $*$   \\ \dotlineb
    & precedences       & \greyed       & \greyed       & \greyed       & \greyed   & \greyed   \\ \dotlineb
    & cluster           & \greyed       & \greyed       & \greyed       & \greyed   & \greyed   \\ \hline
\multirow{4}{1em}{$\delta$}  
    & domain/range      & $c: E \mapsto \mathbb{R}$    & $c: E \mapsto \mathbb{R}$ 
    & $c: E \mapsto \mathbb{R}$   &  $c: E \mapsto \mathbb{R}$ &  $c: E \mapsto \mathbb{R}$ \\ 
    &                   & $p: V \mapsto \mathbb{R}_{\geq 0}$    & &     $p: V \mapsto \mathbb{R}_{\geq 0}$             
    & & $w: E \mapsto \mathbb{R}_{\geq 0}$\\
    &   & $q: V \mapsto \mathbb{R}_{\geq 0} $ & $q: V \mapsto \mathbb{R}_{\geq 0}$ & 
    & $q: V \mapsto \mathbb{R}_{\geq 0}$ & $q: V \mapsto \mathbb{R}_{\geq 0}$\\
     &&&&&& $h: V \mapsto \mathbb{R}_{\geq 0}$\\  \dotlineb
        
    & property          & $*$           & $*$           &   $*$         &  $*$   & $*$\\ \dotlineb
    & partial           & \greyed       & \greyed       & \greyed       & \greyed & \greyed \\ \dotlineb
    & temporal          & \greyed       & \greyed       & \greyed       &  \greyed & $w:$ waiting\\ \hline
$\epsilon$ 
    & objective         & $q(\sol) \geq b; \min c(\sol)+ \compl{p}(\sol)$  
    & $q(\sol) \geq b; \min c(\sol)$                 & $\min c(\sol) + \compl{p}(\sol)$ & $\max q(\sol)$ & $\max q(\sol)$ ; \\ 
    &&&&&& $\forall \, i \in \{0, \dots, k\}: r(v_i) \leq$ \\
    &&&&&& $c(\sol_{<i}) + w(\sol_{\leq i}) + h(\sol_{<i}) \leq$ \\
    &&&&&& $d(v_i) - h(v_i)$  \\ \hline
    \end{tabular}
\end{sidewaystable}

\end{appendices}

\end{document}